\documentclass[10pt, journal]{IEEEtran}
\IEEEoverridecommandlockouts
\usepackage{cite}
\usepackage[most]{tcolorbox}
\usepackage{amsmath}
\usepackage{amsthm}
\usepackage{url}
\usepackage{tikz}
\usepackage{multirow}

\usepackage{colortbl}

\usepackage{graphicx}
\usepackage{xcolor}
\usepackage{color}
\usepackage[normalem]{ulem}

\usepackage{varwidth}
\usepackage{rotating}

\usepackage[nolist]{acronym}

\usepackage{nameref}
\usepackage{zref-xr,zref-user}
\zxrsetup{toltxlabel=true, tozreflabel=false}
\usepackage{xcite}

\usepackage{tikz-3dplot}

\usepackage{subfigure}

\usetikzlibrary{decorations.pathreplacing,decorations.markings,decorations.text}

\tikzset{arrow data/.style 2 args={%
      decoration={%
         markings,
         mark=at position #1 with \arrow{#2}},
         postaction=decorate}
      }%

\usetikzlibrary{shapes,backgrounds,calc}

\tikzstyle{dummy} = [rectangle, text width=0.1em, draw=white, white,
                      minimum width=0.1em, minimum height=3em, opacity=0.0]

\makeatletter
\newcommand*{\Strut}[1][0.1em]{\vrule\@width\z@\@height#1\@depth\z@\relax}
\makeatother

\definecolor{Linen}{rgb}{0.9803,0.9411,0.9019}
\definecolor{White}{rgb}{1,1,1}
\definecolor{Green}{rgb}{0.5,1,0.5}
\definecolor{Red}{rgb}{1,0.4,0.4}
\definecolor{Coral}{rgb}{1,0.4980,0.3137}
\definecolor{Grayblue}{rgb}{0.9411,0.9411,0.9803}
\definecolor{DarkLinen}{rgb}{0.729,0.7176,0.635}

\begin{document}

\begin{acronym}

\acro{5G-NR}{5G New Radio}
\acro{3GPP}{3rd Generation Partnership Project}
\acro{AC}{address coding}
\acro{ACF}{autocorrelation function}
\acro{ACR}{autocorrelation receiver}
\acro{ADC}{analog-to-digital converter}
\acrodef{aic}[AIC]{Analog-to-Information Converter}     
\acro{AIC}[AIC]{Akaike information criterion}
\acro{aric}[ARIC]{asymmetric restricted isometry constant}
\acro{arip}[ARIP]{asymmetric restricted isometry property}

\acro{ARQ}{automatic repeat request}
\acro{AUB}{asymptotic union bound}
\acrodef{awgn}[AWGN]{Additive White Gaussian Noise}     
\acro{AWGN}{additive white Gaussian noise}

\acro{APSK}[PSK]{asymmetric PSK} 

\acro{waric}[AWRICs]{asymmetric weak restricted isometry constants}
\acro{warip}[AWRIP]{asymmetric weak restricted isometry property}
\acro{BCH}{Bose, Chaudhuri, and Hocquenghem}        
\acro{BCHC}[BCHSC]{BCH based source coding}
\acro{BEP}{bit error probability}
\acro{BFC}{block fading channel}
\acro{BG}[BG]{Bernoulli-Gaussian}
\acro{BGG}{Bernoulli-Generalized Gaussian}
\acro{BPAM}{binary pulse amplitude modulation}
\acro{BPDN}{Basis Pursuit Denoising}
\acro{BPPM}{binary pulse position modulation}
\acro{BPSK}{binary phase shift keying}
\acro{BPZF}{bandpass zonal filter}
\acro{BSC}{binary symmetric channels}              
\acro{BU}[BU]{Bernoulli-uniform}
\acro{BER}{bit error rate}
\acro{BS}{base station}

\acro{CP}{Cyclic Prefix}
\acrodef{cdf}[CDF]{cumulative distribution function}   
\acro{CDF}{cumulative distribution function}
\acrodef{c.d.f.}[CDF]{cumulative distribution function}
\acro{CCDF}{complementary cumulative distribution function}
\acrodef{ccdf}[CCDF]{complementary CDF}               
\acrodef{c.c.d.f.}[CCDF]{complementary cumulative distribution function}
\acro{CD}{cooperative diversity}

\acro{CDMA}{Code Division Multiple Access}
\acro{ch.f.}{characteristic function}
\acro{CIR}{channel impulse response}
\acro{cosamp}[CoSaMP]{compressive sampling matching pursuit}
\acro{CR}{cognitive radio}
\acro{cs}[CS]{compressed sensing}                   
\acrodef{cscapital}[CS]{Compressed sensing} 
\acrodef{CS}[CS]{compressed sensing}
\acro{CSI}{channel state information}
\acro{CCSDS}{consultative committee for space data systems}
\acro{CC}{convolutional coding}
\acro{Covid19}[COVID-19]{Coronavirus disease}
\acro{CAPEX}{CAPital EXpenditures}

\acro{DAA}{detect and avoid}
\acro{DAB}{digital audio broadcasting}
\acro{DCT}{discrete cosine transform}
\acro{dft}[DFT]{discrete Fourier transform}
\acro{DR}{distortion-rate}
\acro{DS}{direct sequence}
\acro{DS-SS}{direct-sequence spread-spectrum}
\acro{DTR}{differential transmitted-reference}
\acro{DVB-H}{digital video broadcasting\,--\,handheld}
\acro{DVB-T}{digital video broadcasting\,--\,terrestrial}
\acro{DL}{downlink}
\acro{DSSS}{Direct Sequence Spread Spectrum}
\acro{DFT-s-OFDM}{Discrete Fourier Transform-spread-Orthogonal Frequency Division Multiplexing}
\acro{DAS}{distributed antenna system}
\acro{DNA}{Deoxyribonucleic Acid}

\acro{EC}{European Commission}
\acro{EED}[EED]{exact eigenvalues distribution}
\acro{EIRP}{Equivalent Isotropically Radiated Power}
\acro{ELP}{equivalent low-pass}
\acro{eMBB}{Enhanced Mobile Broadband}
\acro{EMF}{Electro-Magnetic Field}
\acro{EU}{European union}
\acro{ELP}{Exposure Limit-based Power}

\acro{FC}[FC]{fusion center}
\acro{FCC}{Federal Communications Commission}
\acro{FEC}{forward error correction}
\acro{FFT}{fast Fourier transform}
\acro{FH}{frequency-hopping}
\acro{FH-SS}{frequency-hopping spread-spectrum}
\acrodef{FS}{Frame synchronization}
\acro{FSsmall}[FS]{frame synchronization}  
\acro{FDMA}{Frequency Division Multiple Access}  
\acro{FSPL}{Free Space Path Loss}

\acro{GA}{Gaussian approximation}
\acro{GF}{Galois field }
\acro{GG}{Generalized-Gaussian}
\acro{GIC}[GIC]{generalized information criterion}
\acro{GLRT}{generalized likelihood ratio test}
\acro{GPS}{Global Positioning System}
\acro{GMSK}{Gaussian minimum shift keying}
\acro{GSMA}{Global System for Mobile communications Association}

\acro{HAP}{high altitude platform}

\acro{IDR}{information distortion-rate}
\acro{IFFT}{inverse fast Fourier transform}
\acro{iht}[IHT]{iterative hard thresholding}
\acro{i.i.d.}{independent, identically distributed}
\acro{IoT}{Internet of Things}                      
\acro{IR}{impulse radio}
\acro{lric}[LRIC]{lower restricted isometry constant}
\acro{lrict}[LRICt]{lower restricted isometry constant threshold}
\acro{ISI}{intersymbol interference}
\acro{ITU}{International Telecommunication Union}
\acro{ICNIRP}{International Commission on Non-Ionizing Radiation Protection}
\acro{IEEE}{Institute of Electrical and Electronics Engineers}
\acro{ICES}{IEEE international committee on electromagnetic safety}
\acro{IEC}{International Electrotechnical Commission}
\acro{IARC}{International Agency on Research on Cancer}
\acro{IS-95}{Interim Standard 95}

\acro{LEO}{low earth orbit}
\acro{LF}{likelihood function}
\acro{LLF}{log-likelihood function}
\acro{LLR}{log-likelihood ratio}
\acro{LLRT}{log-likelihood ratio test}
\acro{LOS}{Line-of-Sight}
\acro{LRT}{likelihood ratio test}
\acro{wlric}[LWRIC]{lower weak restricted isometry constant}
\acro{wlrict}[LWRICt]{LWRIC threshold}
\acro{LPWAN}{low power wide area network}
\acro{LoRaWAN}{Low power long Range Wide Area Network}
\acro{NLOS}{non-line-of-sight}
\acro{LB}{Lower Bound}

\acro{MB}{multiband}
\acro{MC}{multicarrier}
\acro{MDS}{mixed distributed source}
\acro{MF}{matched filter}
\acro{m.g.f.}{moment generating function}
\acro{MI}{mutual information}
\acro{MIMO}{multiple-input multiple-output}
\acro{MISO}{multiple-input single-output}
\acrodef{maxs}[MJSO]{maximum joint support cardinality}                       
\acro{ML}[ML]{maximum likelihood}
\acro{MMSE}{minimum mean-square error}
\acro{MMV}{multiple measurement vectors}
\acrodef{MOS}{model order selection}
\acro{M-PSK}[${M}$-PSK]{$M$-ary phase shift keying}                       
\acro{M-APSK}[${M}$-PSK]{$M$-ary asymmetric PSK} 
\acro{MSP}{Minimum Sensitivity-based Power}

\acro{M-QAM}[$M$-QAM]{$M$-ary quadrature amplitude modulation}
\acro{MRC}{maximal ratio combiner}                  
\acro{maxs}[MSO]{maximum sparsity order}                                      
\acro{M2M}{machine to machine}                                                
\acro{MUI}{multi-user interference}
\acro{mMTC}{massive Machine Type Communications}      
\acro{mm-Wave}{millimeter-wave}
\acro{MP}{mobile phone}
\acro{MPE}{maximum permissible exposure}
\acro{MAC}{media access control}
\acro{NB}{narrowband}
\acro{NBI}{narrowband interference}
\acro{NLA}{nonlinear sparse approximation}
\acro{NLOS}{Non-Line of Sight}
\acro{NTIA}{National Telecommunications and Information Administration}
\acro{NTP}{National Toxicology Program}
\acro{NHS}{National Health Service}

\acro{OC}{optimum combining}                             
\acro{OC}{optimum combining}
\acro{ODE}{operational distortion-energy}
\acro{ODR}{operational distortion-rate}
\acro{OFDM}{orthogonal frequency-division multiplexing}
\acro{omp}[OMP]{orthogonal matching pursuit}
\acro{OSMP}[OSMP]{orthogonal subspace matching pursuit}
\acro{OQAM}{offset quadrature amplitude modulation}
\acro{OQPSK}{offset QPSK}
\acro{OFDMA}{Orthogonal Frequency-division Multiple Access}
\acro{OPEX}{OPerating EXpenditures}
\acro{OQPSK/PM}{OQPSK with phase modulation}

\acro{PAM}{pulse amplitude modulation}
\acro{PAR}{peak-to-average ratio}
\acrodef{pdf}[PDF]{probability density function}                      
\acro{PDF}{probability density function}
\acrodef{p.d.f.}[PDF]{probability distribution function}
\acro{PDP}{power dispersion profile}
\acro{PMF}{probability mass function}                             
\acrodef{p.m.f.}[PMF]{probability mass function}
\acro{PN}{pseudo-noise}
\acro{PPM}{pulse position modulation}
\acro{PRake}{Partial Rake}
\acro{PSD}{power spectral density}
\acro{PSEP}{pairwise synchronization error probability}
\acro{PSK}{phase shift keying}
\acro{PD}{Power Density}
\acro{8-PSK}[$8$-PSK]{$8$-phase shift keying}

\acro{FSK}{frequency shift keying}

\acro{QAM}{Quadrature Amplitude Modulation}
\acro{QPSK}{quadrature phase shift keying}
\acro{OQPSK/PM}{OQPSK with phase modulator }

\acro{RD}[RD]{raw data}
\acro{RDL}{"random data limit"}
\acro{ric}[RIC]{restricted isometry constant}
\acro{rict}[RICt]{restricted isometry constant threshold}
\acro{rip}[RIP]{restricted isometry property}
\acro{ROC}{receiver operating characteristic}
\acro{rq}[RQ]{Raleigh quotient}
\acro{RS}[RS]{Reed-Solomon}
\acro{RSC}[RSSC]{RS based source coding}
\acro{RFP}{Radio Frequency ``Pollution''}
\acro{r.v.}{random variable}                               
\acro{R.V.}{random vector}
\acro{RMS}{root mean square}
\acro{RFR}{radiofrequency radiation}
\acro{RIS}{Reconfigurable Intelligent Surface}
\acro{RNA}{RiboNucleic Acid}

\acro{SA}[SA-Music]{subspace-augmented MUSIC with OSMP}
\acro{SCBSES}[SCBSES]{Source Compression Based Syndrome Encoding Scheme}
\acro{SCM}{sample covariance matrix}
\acro{SEP}{symbol error probability}
\acro{SG}[SG]{sparse-land Gaussian model}
\acro{SIMO}{single-input multiple-output}
\acro{SINR}{Signal-to-Interference plus Noise Ratio}
\acro{SIR}{signal-to-interference ratio}
\acro{SISO}{single-input single-output}
\acro{SMV}{single measurement vector}
\acro{SNR}[\textrm{SNR}]{signal-to-noise ratio} 
\acro{sp}[SP]{subspace pursuit}
\acro{SS}{spread spectrum}
\acro{SW}{sync word}
\acro{SAR}{Specific Absorption Rate}
\acro{SSB}{synchronization signal block}

\acro{TH}{time-hopping}
\acro{ToA}{time-of-arrival}
\acro{TR}{transmitted-reference}
\acro{TW}{Tracy-Widom}
\acro{TWDT}{TW Distribution Tail}
\acro{TCM}{trellis coded modulation}
\acro{TDD}{time-division duplexing}
\acro{TDMA}{Time Division Multiple Access}

\acro{UAV}{unmanned aerial vehicle}
\acro{uric}[URIC]{upper restricted isometry constant}
\acro{urict}[URICt]{upper restricted isometry constant threshold}
\acro{UWB}{ultrawide band}
\acro{UWBcap}[UWB]{Ultrawide band}   
\acro{URLLC}{Ultra Reliable Low Latency Communications}
         
\acro{wuric}[UWRIC]{upper weak restricted isometry constant}
\acro{wurict}[UWRICt]{UWRIC threshold}                
\acro{UE}{user equipment}
\acro{UL}{uplink}
\acro{UB}{Upper Bound}

\acro{WiM}[WiM]{weigh-in-motion}
\acro{WLAN}{wireless local area network}
\acro{wm}[WM]{Wishart matrix}                               
\acroplural{wm}[WM]{Wishart matrices}
\acro{WMAN}{wireless metropolitan area network}
\acro{WPAN}{wireless personal area network}
\acro{wric}[WRIC]{weak restricted isometry constant}
\acro{wrict}[WRICt]{weak restricted isometry constant thresholds}
\acro{wrip}[WRIP]{weak restricted isometry property}
\acro{WSN}{wireless sensor network}                        
\acro{WSS}{wide-sense stationary}
\acro{WHO}{World Health Organization}
\acro{Wi-Fi}{wireless fidelity}

\acro{sss}[SpaSoSEnc]{sparse source syndrome encoding}

\acro{VLC}{visible light communication}
\acro{VPN}{virtual private network} 
\acro{RF}{Radio-Frequency}
\acro{FSO}{free space optics}
\acro{IoST}{Internet of space things}

\acro{GSM}{Global System for Mobile Communications}
\acro{2G}{second-generation cellular network}
\acro{3G}{third-generation cellular network}
\acro{4G}{fourth-generation cellular network}
\acro{5G}{5th-generation cellular network}	
\acro{gNB}{next-generation Node-B}
\acro{NR}{New Radio}
\acro{UMTS}{Universal Mobile Telecommunications Service}
\acro{LTE}{Long Term Evolution}
\acro{SPS}{Spectrum-based Power Setting}

\acro{QoS}{Quality of Service}
\end{acronym}

\title{``{Cellular Network} Densification Increases Radio-Frequency Pollution'': True or False?\footnote{A preliminary version of this work has been presented at the IEEE VTC-Spring 2020 conference \cite{chiaraviglio2020will}.}}
\author{\small  Luca Chiaraviglio,$^{(1,2)}$ Sara Turco,$^{(1,2)}$ Giuseppe Bianchi,$^{(1,2)}$, Nicola Blefari-Melazzi,$^{(1,2)}$\\
(1) Department of Electronic Engineering, \\University of Rome Tor Vergata, Rome, Italy, email \{luca.chiaraviglio,giuseppe.bianchi,blefari\}@uniroma2.it\\
(2) Consorzio Nazionale Interuniversitario per le Telecomunicazioni, Italy, email sara.turco@cnit.it
}
\maketitle
\IEEEpeerreviewmaketitle
\vspace{-2cm}
\begin{abstract}
A very popular theory circulating among non-scientific communities claims that the massive deployment of  {Base Stations (BSs)} over the territory, a.k.a.  {cellular network} densification, always triggers an uncontrolled and exponential increase of human exposure to \ac{RFP}. To face such concern in a way that can be understood by the layman, in this work we develop a very simple model to compute the \ac{RFP}, based on a set of worst-case and conservative assumptions. We then provide closed-form expressions to evaluate the \ac{RFP} variation in a pair of candidate 5G deployments, subject to different densification levels. Results, obtained over a wide set of representative 5G scenarios, dispel the myth:  {cellular network} densification triggers an \ac{RFP} decrease {(up to three orders of magnitude)} when the radiated power from  {the BS} is adjusted to ensure a minimum sensitivity at the cell edge. Eventually, we analyze the conditions under which the \ac{RFP} may increase when the network is densified (e.g., when the radiated power does not scale with the cell size), proving that the amount of \ac{RFP} is always controlled. Finally, the results obtained by simulation confirm the outcomes of the RFP model.
\end{abstract}


\section{Introduction}
\label{sec:intro}

Unlike past generations cellular systems, the ongoing deployment of 5G networks is surprisingly raising severe debates and strong concerns among the population, to the extent that 5G is sometimes even perceived as a threat.  {In general}, the installation of  {\acp{BS}} over the territory  {(called \acp{gNB} in 5G)} often generates a sentiment of suspect and/or fear, since \ac{RF} exposure from  {\acp{BS}} is connected to the emergence of severe health effects \cite{healthconcerns}. Although the causal correlation between \ac{RF} exposure below the limits defined by laws and long-term effects over humans has not been scientifically (and widely) proven so far by the research community (see e.g., the recent surveys \cite{simko20195g,bushberg2020ieee}), the continuous fabrication of myths and false claims about health effects due to  {\ac{BS}} exposure fuels a diffuse negative feeling against this technology \cite{5Grisks}, leading to sabotages of towers hosting cellular equipment \cite{sabotage3}, and even 5G installation bans promoted by countries/municipalities \cite{pngban,jaimaicaban,brusselsban,californiaban}. 


In this scenario, part the population is firmly convinced that the installation of a massive number of  {\acp{BS}} over the territory - a step often known as  {cellular network} densification - results into an uncontrolled and unacceptable increase of human exposure to Radio Frequency ``Pollution'' (RFP).\footnote{The term``pollution'' is intentionally left inside quotation marks because it is commonly used by the population rather than by the scientific community, who generally adopts more neutral terms like exposure, radiation, and emission.} 
 {Actually,} the underlying intuitive (sic!) layman argument is that the \ac{RFP} directly depends on the amount of deployed antennas: the greater the {number} of \ac{EMF} sources (in terms of  {\acp{BS}}), the greater and more dangerous the relevant \ac{EMF} exposure. And even if any student or practitioner in telecommunications engineering would readily spot the bias in this argument (as the power emitted by  {\acp{BS}} is not an \textit{a-priori} fixed parameter), announcements about dramatical increases of \ac{RFP} due to  {cellular network} densification are frequently spreading across social networks and on {newspapers} \cite{increasedenergy}. 


In this context, two natural questions emerge, namely: How does  {cellular network} densification influence \ac{RFP}? And is the alleged \ac{RFP} increase due to  {\ac{BS}} densification corroborated by scientific evidence? Our goal is to provide an answer to these intriguing questions. Although the scientific community well knows that  {cellular network} densification does not lead in general to an uncontrolled increase of \ac{RFP}, to the best of our knowledge, there is a huge gap between the research outcomes on one side and their comprehension level by the population on the other one. In more detail, answering to the aforementioned questions in a way that can be understood by the layman is still an underrated and relatively neglected aspect so far. To face this gap, in this work we develop a very simple mathematical model to assess the \ac{RFP} increase/decrease when comparing two candidate 5G deployments (e.g., a sparse set  vs. a dense one). Our model, which can be potentially understood even by the general public (with basic mathematical skills), is based on a set of simplifying (but worst-case) assumptions that allow us to derive closed-form expressions for the \ac{RFP}, given as input parameters the main wireless features that characterize a 5G deployment, e.g., the adopted frequency, the propagation conditions, the coverage size of the cell, the deployment tessellation and the setting for the maximum radiated power of the cells. By comparing the \ac{RFP} across pairs of candidate 5G deployments, we are able to assess the impact of  {cellular network} densification and consequently to provide an answer based on scientific evidence to the population concerns.

In order to derive a meaningful set of results, we consider two extreme - yet meaningful - rules to set the radiated power for each \ac{gNB}, denoted as \ac{MSP} and \ac{ELP}, respectively. With \ac{MSP}, the radiated power ensures a minimum sensitivity threshold at the  {\ac{gNB}} edge (and therefore it scales with the  {\ac{BS} coverage area}). With \ac{ELP}, the radiated power is set to ensure a stringent \ac{EMF} limit enforced by law (and therefore it does not scale with the  {\ac{BS} coverage area}). Our outcomes demonstrate that  {cellular network} densification does not increase the \ac{RFP} when \ac{MSP} is adopted. On the other hand, we show that \ac{ELP} may actually increase the \ac{RFP} as the network is densified. However, there are conditions under which the \ac{RFP} is decreased even with this policy, e.g., when the  {\ac{BS}} operating frequency is increased. In all cases, however, the \ac{RFP} variation is always controlled. Eventually, we show that the outcomes of our \ac{RFP} model are confirmed by the numerical values obtained by simulation.  

\subsection{ {Limitations of our work}}

{Clearly, the scope of our work is not to provide an omni-comprehensive evaluation of pollution from  {cellular networks}, but rather to focus on a specific feature: the proliferation of  {\acp{BS}} of the same type over the territory, i.e., a step normally referred as \textit{horizontal} densification} \cite{chiaraviglio2020health}{. Other  {specific technology features}, which include  {5G functionalities like} dynamic beamforming,  {massive \ac{MIMO}} and/or multiple layers of \acp{gNB} covering the same area of territory (a.k.a. \textit{vertical} densification), may introduce a variation in the \ac{RFP} levels w.r.t. the results presented in this work. To this aim, preliminary results in} \cite{chiaraviglio2021pencil} {suggest that densification may be effective in reducing the \ac{RFP} of beamforming  {in 5G networks}, mainly because the propagation conditions between the serving \ac{gNB} and the user are improved. However, a deeper evaluation of this aspect is left for future work.  

\subsection{ {Work Organization}}

The rest of the paper is organized as follows. Sec.~\ref{sec:rel_work} reports the positioning of our work w.r.t. the literature. Sec.~\ref{sec:sys_model} presents the \ac{RFP} model. Sec.~\ref{sec:scenario_definition} details the considered 5G scenarios. Sec.~\ref{sec:evaluation} evaluates the \ac{RFP} over the different 5G scenarios. Finally, Sec.~\ref{sec:conclusions} summarizes the paper and points out possible future activities.

\section{Related Works}
\label{sec:rel_work}



In terms of related works, we could not find papers similar to ours, i.e., specifically focused on the relation between  {cellular} network densification and \ac{RFP}. For this reason, in what follows, we position our work w.r.t. the relevant literature appeared in the following related areas: \textit{i}) performance assessment of 5G densification, \textit{ii}) network planning of dense 5G networks, and \textit{iii}) exposure concerns of 5G networks.

\subsection{Performance Assessment of 5G Densification}

The works falling inside this category \cite{thurfjell2015network, kountouris2017performance,liu2017network,ge20165g,park2014asymptotic,arshad2016handover,andrews2016we,shafi20175g} aim at evaluating the impact of 5G densification on the network performance. To this purpose, Thurfjell \textit{et al.} \cite{thurfjell2015network} demonstrate that the user bit rate tends to improve when the network is densified. However, the benefits in terms of capacity may be negatively impacted by the level of interference from the neighboring cells, as well as changes in the path loss exponents, as pointed out by the same authors. The performance limits of network densification are further analyzed by Nguyen and Kountouris, \cite{kountouris2017performance}, demonstrating that the user performance increases up to a certain level of densification, after which a saturation or even decay with increasing network density is observed.  Moreover, the asymptotic behavior of spectral efficiency is faced by Park \textit{et al.} \cite{park2014asymptotic}, showing that this metric always grows with increased densification levels. According to Andrews \textit{et al.} \cite{andrews2016we}, the \ac{SINR} initially increases as the 5G network is densified and then decreases after a given level of 5G densification.

Other effects triggered by 5G densification include constraints on the required backhaul network capacity \cite{ge20165g}, large requirements in terms of backhaul energy consumption \cite{ge20165g} and increasing handover rates \cite{arshad2016handover}. Moreover, possible approaches to enhance network capacity are analyzed by Liu \textit{et al.} \cite{liu2017network}, due to the fact that short-range propagation conditions tend to change in ultra-dense networks w.r.t. legacy ones. Eventually, the importance of introducing interference cancellation mechanisms as network is densified is pointed out by Shafi \textit{et al.} \cite{shafi20175g}. Finally, densification will be an important feature also in beyond-5G cellular networks, as claimed by Dang \textit{et al.} \cite{dang2020should}.

Summarizing, the performance gains and the fundamental limits of 5G densification are thoroughly analyzed by the related literature. Despite we recognize the importance of such previous works, none of them investigate the impact of densification on the \ac{RFP}, which is a major concern for the population and the main goal of this paper.

\subsection{Planning of 5G Dense Cellular Networks}

Works \cite{oughton2019open,chiaraviglio2018planning,aaltod3} focus on the design of 5G dense cellular networks under costs, coverage and regulatory constraints. The overall problem, often known as 5G network planning, aims at minimizing \ac{CAPEX} and \ac{OPEX} costs for the installed \acp{gNB}, as well as at properly configuring each installed site in terms of, e.g., radiating elements, antenna configurations, maximum radiated power from each \ac{RF} element, etc. Clearly, the \ac{CAPEX} costs tend to notably increase when the inter-site distance is reduced (and consequently the densification level is increased), as shown by Oughton \textit{et al.} \cite{oughton2019open}. In addition, the planning of dense cellular networks is severely limited in countries ensuring strict \ac{EMF} limits \cite{chiaraviglio2018planning}, which prevent the installation of new \acp{gNB} over the territory, due to the fact that the overall exposure levels from legacy technologies (e.g., radio/TV repeaters, 2G/3G/4G Base Stations) are already close to the maximum \ac{EMF} limits. Eventually, the importance of regulatory updates to support the planning of dense 5G networks is stressed by \cite{aaltod3}. 

Although we recognize the importance of the 5G planning problem, only a subset of previous works (e.g., \cite{chiaraviglio2018planning,aaltod3}) consider \ac{EMF} constraints, without however evaluating the impact of different densification levels on the \ac{RFP}. In contrast to them, our goal is to provide a simple - yet effective -  model to compute the \ac{RFP} level of a  {cellular} network, and to evaluate the variation on the \ac{RFP} when the network is densified. Clearly, the  \ac{RFP} contributions from legacy technologies are intentionally not treated in this work, since our goal is to assess the \ac{RFP} from 5G  {deployments}.

\subsection{Exposure Concerns from 5G Networks}

A third group of works \cite{simko20195g,bushberg2020ieee,colombi2020analysis} is instead tailored to the analysis and assessment of exposure concerns from 5G networks. Simk{\'o} and Mattsson \cite{simko20195g} review the related literature about health effects from 5G (and pre-5G) exposure, concluding that there is not a consistent relation between health effects and exposure levels, exposure durations or frequency. However, the authors point out that a meaningful safety assessment can not be retrieved from the available studies, and so further researches are needed e.g., to thoroughly assess the (possible) health implications of non-thermal effects triggered by 5G exposure. The impact of 5G on the levels of exposure is also discussed by the \ac{IEEE} Committee on Man and Radiation in \cite{bushberg2020ieee}. In particular, the committee members point out that 5G densification will increase the downlink signal levels, which in turn may reduce the radiated power in the uplink radiation, and hence the exposure from terminals. In addition,  the exposure levels will remain lower than the maximum limits ensured by the regulations (based on the guidelines promoted by international organizations such as \ac{IEEE} \cite{IEEEC95:19} and \ac{ICNIRP} \cite{ICNIRPGuidelines:20}), even when the network is densified  \cite{bushberg2020ieee}. Finally, Colombi \textit{et al.} \cite{colombi2020analysis} perform an exposure assessment in a commercial 5G network, demonstrating that the maximum time-averaged power per beam direction is notably lower than the theoretical maximum. Consequently, the concerns associated with very high exposure levels generated by 5G \acp{gNB} are not justified in practice.

In contrast to these works, we do not directly address the relation between \ac{EMF} exposure and health. Rather, we aim at \textit{quantifying} the \ac{RFP} when the  {cellular} network is densified. In particular, we demonstrate that there are conditions under which  {cellular network} densification triggers an \ac{RFP} reduction, which is in turn beneficial in alleviating the exposure concerns.

\section{Radio Frequency ``Pollution'' Model}
\label{sec:sys_model}

In this section, we describe the main building blocks that characterize our \ac{RFP} model, and namely: \textit{i}) main assumptions, \textit{ii}) \ac{RFP} definition,  \textit{iii}) radiated power setting, \textit{iv}) cell \ac{RFP} model, \textit{v}) \ac{RFP} model at fixed distance, \textit{vi}) \ac{RFP} upper bound from neighbors, and \textit{vii)} \ac{RFP} ratio among 5G deployments. 

\subsection{Main Assumptions}

Our model leverages some standard topological/regularity/propagation assumptions, namely: 
\begin{enumerate} 
\item The \acp{gNB} are placed on a regular layout, as we consider a dense {urban} deployment with a uniform distribution of users; consequently, each \ac{gNB} serves a portion of the total territory under consideration. {This assumption is inline with the 5G scenarios defined by relevant standardization bodies (such as \ac{3GPP}), which adopt regular deployments and regular cell layouts for urban case-studies} \cite{3GPPscenarios};\footnote{{Clearly, other scenarios (including e.g., a mixture of urban/rural areas and/or non-uniform distribution of users) may require a different modeling for the positions of \acp{gNB} and/or users, based e.g., on stochastic geometry tools. We leave the investigation of such aspects as future work.}}
\item All the \acp{gNB} of a given deployment are characterized by common features in terms of coverage shape, coverage size, maximum radiated power and adopted frequency; i.e., the same \ac{gNB} equipment is used across the set. {This assumption is motivated by the fact that our goal is to evaluate the proliferation of the same type of \acp{gNB}, i.e., the horizontal densification.}\footnote{{The investigation of the impact of heterogeneous 5G networks (i.e., composed of multiple layers of \acp{gNB} simultaneously providing coverage over the same area) is left for future work.}} 
\item The propagation conditions are the same among the \acp{gNB} in the set; e.g., the reliable coverage distance is sufficiently short to avoid modifications of the propagation model due to changes in the path loss exponent \cite{rappaport2017overview}.\footnote{The integration of more complex propagation models, e.g., based on a dual slope, is left for future work.} {To this aim, we employ coverage distances in the ranges specified by 3GPP for urban deployments} \cite{3GPPscenarios}.
\end{enumerate}

In addition, another key feature that is assumed in this work is an omnidirectional pattern to characterize the \ac{gNB} radiation. Clearly, a real 5G \ac{gNB} generally exhibits a radiation pattern different than a omnidirectional one, because: \textit{i}) sectorization is in general exploited, and \textit{ii}) the extensive adoption of beamforming allows concentrating the transmitted signal strength on specific territory locations. With sectorization, the radiation patterns match the orientation of the sectors. With beamforming, the actual \ac{RFP} level that is received over the territory generally varies both in time and space, and it is normally estimated through statistical models, which demonstrate that the average \ac{RFP} at a given pixel is substantially lower than the theoretical maximum value \cite{Tho-17,colombi2020analysis}. In our case, assuming an omnidirectional radiation is a worst case scenario, in which: \textit{i}) each pixel of the territory is served by a beam (i.e., the beams are simultaneously activated in all the directions), \textit{ii}) each pixel is not affected by sectorization (i.e., for a given pixel to \ac{gNB} distance, the \ac{UE} received power is constant across the entire geographic extent of the sector, even for pixels along the sector edge).

{In more detail, the omni-directional assumption leads to an over-estimation of the received \ac{RFP}, which substantiates our results. Let us provide more explanations about this aspect, by first assuming that not all beams are activated all together at the same time. In this case, a subset of beams is activated to cover only the zones where the users are currently located. Interestingly, our model correctly estimates the \ac{RFP} for the zones exposed to the active beams. On the other hand, the zones that are not served by the beams are subject to a negligible amount of pollution. Therefore, in this scenario, our conclusions are valid for the pollution received by the served users. In other cases, however, the \ac{RFP} may be higher than the one estimated in this work. Such additional scenarios include: \textit{i}) multiple beams simultaneously serving multiple users over the same portion of territory and/or \textit{ii}) each user simultaneously served by more than one beam. Despite we recognize the importance of scenarios \textit{i})-\textit{ii}), we point out that in many countries the currently adopted option to deploy 5G \acp{gNB} is to use static and non overlapping beams (see e.g., the recent work of} \cite{franci2020experimental}{). Moreover, power lock mechanisms are under study to turn off the beam(s) if the overall exposure is excessive} \cite{adda2020methodology}{. Eventually, preliminary results in} \cite{chiaraviglio2021pencil} {demonstrate that densification is beneficial in reducing the exposure in networks applying dynamic beamforming, mainly because the propagation conditions between the serving \ac{gNB} and the user are improved, thus yielding to a decrease of the power radiated by the beam.}

\subsection{RFP Definition}

In this work, the \ac{RFP} is defined as the amount of power that is received over a given pixel $p$ from the serving \ac{gNB} $s$ \textit{and} from each \ac{gNB} $i$ in the neighborhood $\mathcal{I}^{\text{NEIGH}}$. Actually, other alternative metrics that can be exploited to characterize \ac{RFP} include \ac{EMF} strength, \ac{PD} and/or \ac{SAR}. We refer the interested reader to \cite{jamshed2019survey} for an overview about the main \ac{RFP} metrics. Henceforth, we consider the received power as the reference metric, due to the following reasons: \textit{i}) we exploit well-known propagation models derived from telecommunications research (see e.g., \cite{rappaport2017overview}) to compute the \ac{RFP} levels over the territory, \textit{ii)} we consider different rules, including a minimum sensitivity threshold at the cell edge, to set the \ac{gNB} radiated power.{}

Let us assume a standard propagation model \cite{rappaport2017overview}, in which the received \ac{RFP} depends on the power radiated by the \acp{gNB}, scaled by the propagation parameters. More formally, the total \ac{RFP} $P^R_{(p)}$ over pixel $p$ is denoted as:
\begin{equation}
\label{eq:gen_model}
P^R_{(p)}=\underbrace{\frac{P^E}{d_{(p,s)}^\gamma \cdot f^\eta \cdot c}}_\text{RFP from serving gNB} + \underbrace{\sum_{i \in \mathcal{I}^{\text{NEIGH}}} \frac{P^E}{d_{(p,i)}^\gamma \cdot f^\eta \cdot c}}_\text{RFP from neighboring gNBs},
\end{equation}
where $P^E$ is the \ac{gNB} emitted power,  $d_{(p,s)}$~[m] is the distance between the serving \ac{gNB} $s$ and the current pixel $p$, $\gamma$ is the propagation exponent for the distance, $f$~[GHz] is the operating frequency, $\eta$ is the frequency exponent, $c$ is a constant integrating other effects (e.g., the fixed term in the Friis' free space equation \cite{friis1946note}), and $d_{(p,i)}$~[m] is the distance between neighboring \ac{gNB} $i$ and the current pixel $p$.

{By observing in more detail Eq.~(\ref{eq:gen_model}), we can note that the \ac{RFP} significantly differs w.r.t. other metrics commonly adopted for energy and/or performance evaluations (e.g., area power efficiency, area spectral efficiency and peak power), which either include the contribution of neighbors as interference or they only consider the power of the serving \ac{gNB}. In our work, the \ac{RFP} is the summation of power from all the \acp{gNB} that contribute to the exposure (i.e., both serving and interfering ones).}

Clearly, a natural question is: How {do we} select the serving \ac{gNB} $s$ and the neighboring \acp{gNB} $i$ for a given pixel $p$ appearing in Eq.~(\ref{eq:gen_model})? To this aim, we assume a circular coverage area of radius $d_{\text{MAX}}$, which corresponds to the maximum coverage distance. In addition, we introduce a minimum distance $d_{\text{MIN}}$ to model the presence of an exclusion zone in proximity to the \ac{gNB}. In line with the international recommendations \cite{iturec} and the exposure assessment standards \cite{iecexclusion}, the exclusion zone access is forbidden to the general public, and therefore this zone is not considered in the \ac{RFP} computation. Consequently, \ac{gNB} $s$ serves pixel $p$ if $d_{\text{MIN}}\leq d_{(p,s)} \leq d_{\text{MAX}}$. Focusing then on the neighboring \acp{gNB}, in this work we assume that the closest $|\mathcal{I}^{\text{NEIGH}}|=N^I$ \acp{gNB} w.r.t. $s$ contribute to the \ac{RFP}. Moreover, we consider different values of  $N^I$ to evaluate its impact on the \ac{RFP}. 

\begin{figure}[t]
\centering
\subfigure[Exact \ac{RFP} Computation]
{
	\includegraphics[width=6cm]{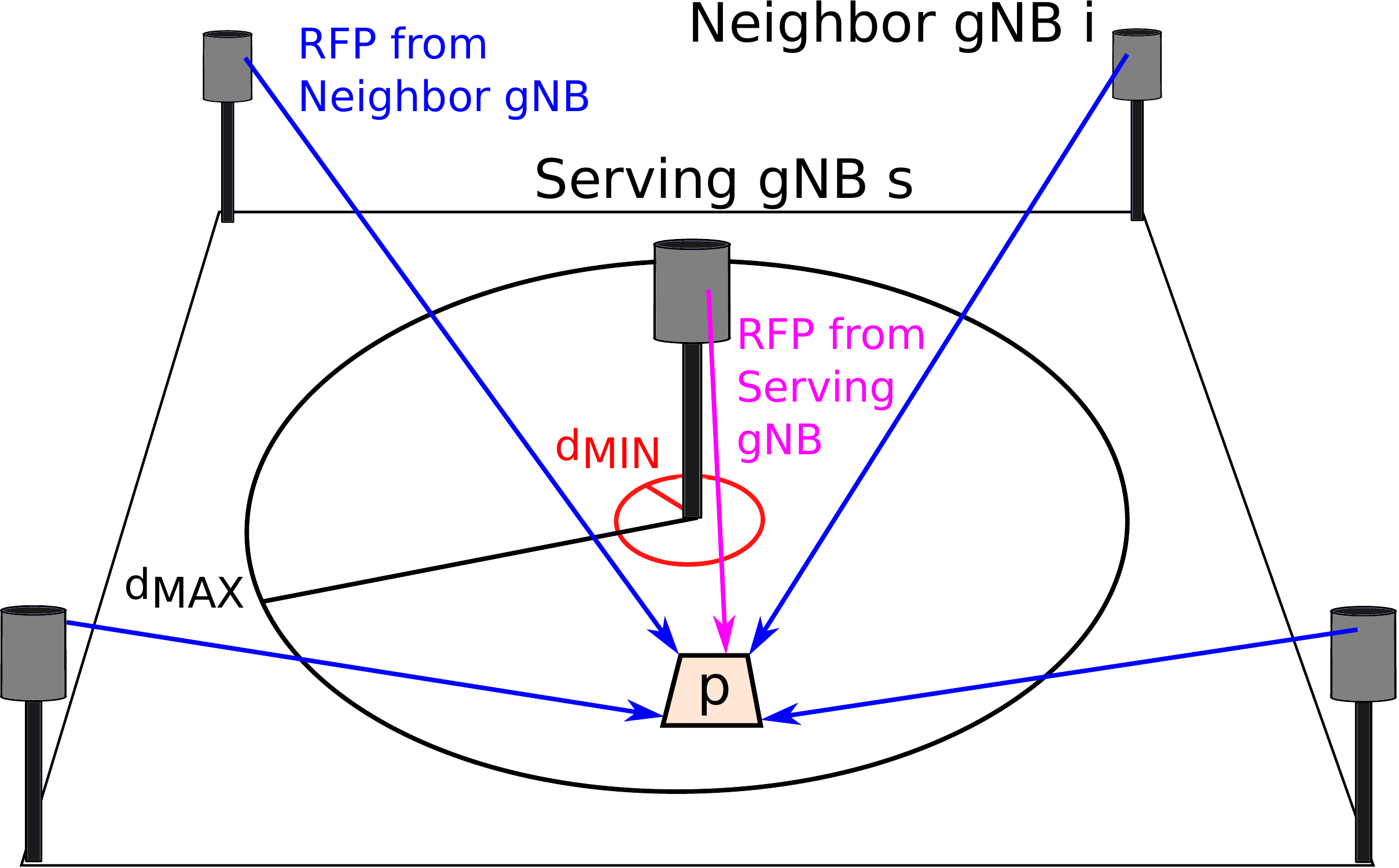}
	\label{fig:rfp_exact_comp}
}
\subfigure[Adopted \ac{RFP} Modepls and Upper Bound]
{
	\includegraphics[width=6cm]{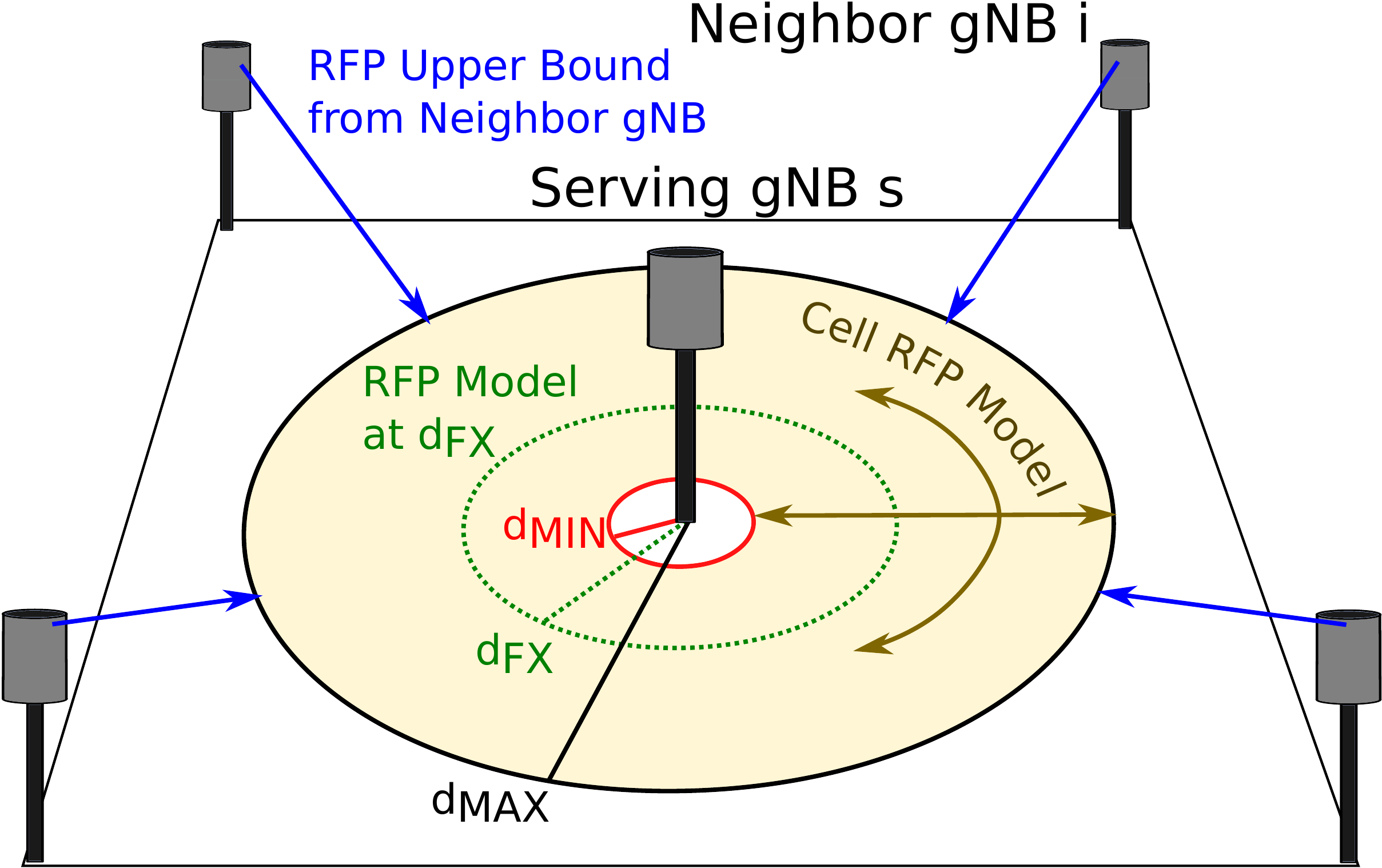}
	\label{fig:rfp_exact_models}
}
\caption{Exact \ac{RFP} computation over a single pixel $p$ (a) vs. adopted \ac{RFP} models and upper bound (b).}
\label{fig:rfp_exact_comp_and_models}
\vspace{-3mm}
\end{figure}

Since our goal is to provide a simple model, we leverage a further worst-case approximation which permits to dramatically simplify computations, and neglects the specific distances $d_{(p,i)}$ from the neighbors appearing in Eq.~(\ref{eq:gen_model}). Such key assumption consists in bounding the ``true'' distance $d_{(p,i)}$ from neighbor $i$ with a {\em constant} (and shorter) distance between each neighbor and the pixel in the target coverage area closer to it, i.e., at the cell edge. This worst-case assumption, which will formally presented as an \ac{UB} in Sec.~\ref{sec:rfp_neigh_cells}, yields a constant and same \ac{RFP} from neighbors for every considered pixel in the coverage area, and allows us to obtain simple closed-form form \ac{RFP} expressions that only depend on $N^I$.

To give more insight, Fig.~\ref{fig:rfp_exact_comp} shows a graphical example of the \ac{RFP} terms appearing in Eq.~(\ref{eq:gen_model}) in a simple toy-case scenario composed of five \acp{gNB}. The considered pixel $p$ falls inside the serving area of \ac{gNB} $s$. Consequently, the \ac{RFP} is computed from the serving \ac{gNB} $s$ and the neighboring $N^I=4$ \acp{gNB}.
Fig.~\ref{fig:rfp_exact_models} then provides a graphical overview on how the \ac{RFP} computation of Eq.~(\ref{eq:gen_model}) is simplified in our models. In particular, we introduce: \textit{i)} a model to compute the \ac{RFP} over the whole area of the serving \ac{gNB}, a.k.a. cell \ac{RFP} model, \textit{ii}) a model to compute the \ac{RFP} at a given distance $d_{\text{FX}}$ from the serving \ac{gNB}, a.k.a. \ac{RFP} model at fixed distance, and \textit{iii)} a \ac{UB} to estimate the \ac{RFP} from neighbors. In parallel, we also build a simulator to compare the outcomes of our models w.r.t. the exact \ac{RFP} computation of Eq.~(\ref{eq:gen_model}).

\subsection{Radiated Power Setting}

Before unveiling the details of our \ac{RFP} models, let us now focus on the radiated power $P^E$ appearing in the \ac{RFP} definition of Eq.~(\ref{eq:gen_model}). Clearly, the value assigned to $P^E$ plays a key role in determining the level of \ac{RF} ``Pollution''. Intuitively, the higher is the radiated power, the higher is also the \ac{RFP} over the territory. Therefore, a careful setting of $P^E$ is crucial for a meaningful \ac{RFP} evaluation. We remind that, in this work, we consider two distinct rules to set $P^E$, namely: \textit{i}) a \ac{MSP} setting and \textit{ii)} an \ac{ELP} one. In the following, we formally describe such policies.

\subsubsection{Minimum Sensitivity-based Power Setting (MSP)}
The goal of \ac{MSP} is to tune $P^E$ in order to ensure a minimum sensitivity threshold $P_{\text{TH}}^R$ at the cell edge $d_{\text{MAX}}$. {We recall that the minimum sensitivity is reported in relevant 5G 3GPP standards} \cite{minsens}{. In particular, $P_{\text{TH}}^R$ is defined as the minimum mean power applied to each one of the \ac{UE} antenna ports in order to meet the throughput requirements. By imposing a minimum sensitivity constraint, we ensure that the throughput is larger than 95\% of the maximum throughput of the reference measurement channels for the users in the worst propagation conditions (i.e., the ones located at the cell edge).}

{More formally, we assume the same propagation model of Eq.~(\ref{eq:gen_model}) to define the radiated power $P^E$:}
\begin{equation}
\label{eq:pe_sens}
  P^E = P_{\text{TH}}^R  \cdot d^{\gamma}_{\text{MAX}}  \cdot f^\eta  \cdot c.
\end{equation}
By replacing Eq.~(\ref{eq:pe_sens}) in Eq.~(\ref{eq:gen_model}) (and without considering the contributions from neighbors), it is trivial to verify that $P^R_{(p)}=P_{\text{TH}}^R$ for the pixel(s) at $d_{\text{MAX}}$.

By observing in more detail Eq.~(\ref{eq:pe_sens}), we can note that the \ac{MSP} setting increases the radiated power when the cell size is widened (i.e., $d_{\text{MAX}}$ increase), when the propagation conditions are worsened (i.e., $\gamma$ increase) and/or when the adopted \ac{gNB} frequency $f$ is increased.

\subsubsection{Exposure Limit-based Power Setting (ELP)} We then consider a second alternative policy for setting $P^E${, which matches the actual power settings in countries imposing strict \ac{EMF} limits} \cite{chiaraviglio2020health}. {When strict \ac{EMF} limits are imposed}, in fact, the power radiated by \acp{gNB} is strongly influenced by the exposure limits for the general public, which have to be ensured in each pixel of the territory (outside the \ac{gNB} exclusion zone) \cite{chiaraviglio2018planning}. Such limits are in general, much more stringent than the international ones \cite{ICNIRPGuidelines:20}, and typically {impose strong limitations on the setting of the} power radiated by each \ac{gNB} \cite{chiaraviglio2018planning}. {In this scenario, when a strict exposure limit has to be enforced, the tightest constraint in network deployment is not performance (like in the \ac{MSP} case), but rather the strict \ac{EMF} exposure limit. To this aim, we refer the interested reader to} \cite{itutks14} {for more details about the impact of strict exposure limits on the performance of mobile networks (including 5G).} 

{More formally, we adopt the procedure defined by \ac{ITU}-T REC K.70} \cite{itutk70} {to verify the adherence w.r.t. the \ac{EMF} limits, by performing the following steps: \textit{i}) we compute the} total \ac{PD} {that is} received by a given pixel from a set of \acp{gNB}, and \textit{ii}) {we verify that} the total received \ac{PD} is lower than the maximum \ac{PD} limit $S_{\text{MAX}}$ for all the pixels outside the \ac{gNB} exclusion zone. Focusing on \textit{i}), we apply the point-source model \cite{itutk70} because the \ac{PD} computed with this approach is always a \ac{UB} of the actual level of exposure, and hence a worst-case scenario. More formally, each \ac{gNB} is characterized by a given transmission gain $G_{\text{TX}}$ and a given transmission loss $L_{\text{TX}}$. The received \ac{PD} $S_{(p,s)}$ by pixel $p$ from \ac{gNB} $s$ is then expressed as in \cite{itutk70}:
\begin{equation}
S_{(p,s)}=\underbrace{\frac{P^E \cdot G_{\text{TX}}}{4\pi \cdot L_{\text{TX}} \cdot d_{(p,s)}^2}}_\text{Point-source model}.
\end{equation}
In order to satisfy the \ac{EMF} limits, the total received \ac{PD} has to be lower than the maximum one:
\begin{equation}
\label{eq:tot_power_density}
\underbrace{\frac{P^E \cdot G_{\text{TX}}}{4\pi \cdot L_{\text{TX}} \cdot d_{(p,s)}^2}}_\text{PD from serving \ac{gNB} $S_{(p,s)}$} + \sum_{i \in \mathcal{I}^{\text{NEIGH}}} \underbrace{\frac{P^E \cdot G_{\text{TX}}}{4\pi \cdot L_{\text{TX}} \cdot  d_{(p,i)}^2}}_\text{PD from neigh. \ac{gNB} $S_{(p,i)}$}
 \leq \underbrace{S_{\text{MAX}}}_\text{PD Limit}
\end{equation}

By observing in more detail Eq.~(\ref{eq:tot_power_density}), we can note that the maximum \ac{PD} is likely experienced in proximity to $s$, where the \ac{PD} contribution from the serving \ac{gNB} $s$ dominates over the one from the neighbors, because $d_{(p,s)}\ll d_{(p,i)}$. This finding is also corroborated by  \ac{EMF} measurements performed over real cellular networks under operation \cite{isitsafe2020}. Therefore, a sufficient condition to satisfy Eq.~(\ref{eq:tot_power_density}) is to verify that the \ac{PD} from the serving \ac{gNB} $s$, evaluated at distance $d_{\text{MIN}}$, is lower than the \ac{PD} limit $S_{\text{MAX}}$:
\begin{equation}
\label{eq:tot_power_density_2}
\underbrace{\frac{P^E \cdot G_{\text{TX}}}{4\pi \cdot L_{\text{TX}} \cdot d_{\text{MIN}}^2}}_\text{PD at minimum distance} \leq S_{\text{MAX}}.
\end{equation}

Consequently, the previous inequality is satisfied when $P^E$ is set equal to:
\begin{equation}
\label{eq:pe_exp_limit}
 P^E = 4 \pi \cdot d_{\text{MIN}}^2 \cdot S_{\text{MAX}} \cdot \frac{L_{\text{TX}}}{G_{\text{TX}}}. 
\end{equation}

By setting $P^E$ in accordance with Eq.~(\ref{eq:pe_exp_limit}), the radiated power does not depend neither on the maximum coverage distance $d_{\text{MAX}}$, the frequency $f$ or the propagation exponent $\gamma$, but solely on the exposure limit  $S_{\text{MAX}}$, the transmission gain $G_{\text{TX}}$, the transmission loss $L_{\text{TX}}$ and the minimum distance $d_{\text{MIN}}$. Clearly, when stringent \ac{EMF} limits are assumed for $S_{\text{MAX}}$, this policy dominates over the \ac{MSP} setting, because the operator always aims at saturating the radiated power to the maximum allowed one.

\subsection{Cell RFP Model} 

In the following, we define a model to capture the \ac{RFP} across the entire cell with a closed-form expression. Our intuition is, in fact, to derive in a compact way the \ac{RFP} from the serving \ac{gNB} $s$ of Eq.~(\ref{eq:gen_model}) for all the pixels $p$ belonging to its coverage area. By assuming an infinitesimal pixel size, we express the average \ac{RFP} over the entire area $\mathcal{A}$ served by the \ac{gNB} as:
\begin{equation}
\label{eq:pavg_fixed}
    P_{\text{CELL}}^R = \frac{1}{\pi(d_{\text{MAX}}^2-d_{\text{MIN}}^2)} \int \int_{\mathcal{A}} \; \frac{P^E}{(\sqrt {x^2+y^2})^\gamma \cdot f^\eta \cdot c} \; \mathrm{d}x \mathrm{d}y,
\end{equation}
where $\frac{1}{\pi(d_{\text{MAX}}^2-d_{\text{MIN}}^2)}$ is the inverse of the served area and \begin{math} \sqrt{x^2+y^2} \end{math} is the distance from the \ac{gNB} center to a generic point \begin{math}(x,y)\end{math}.

In order to solve Eq.~(\ref{eq:pe_sens}), we adopt a reference system based on polar coordinates, where: $x=r\cos{\theta}$, $y=r\sin{\theta}$, $\mathrm{d}x\mathrm{d}y=r\mathrm{d}r\mathrm{d}\theta$. Therefore, Eq.~(\ref{eq:pavg_fixed}) is rewritten as: 
\begin{equation}
\label{eq:pavg_fixed_polar}
    P_{\text{CELL}}^R =\frac{1}{\pi(d_{\text{MAX}}^2-d_{\text{MIN}}^2)} \int_{0}^{2\pi} \int_{d_{\text{MIN}}}^{d_{\text{MAX}}} \frac{P^E}{r^{(\gamma-1)} \cdot f^\eta \cdot c} \mathrm{d}r\mathrm{d}\theta.
\end{equation}

We then consider the solution of Eq.~(\ref{eq:pavg_fixed_polar}) for $\gamma \in (2,4]$, i.e., a typical range for characterizing \ac{LOS}/\ac{NLOS} conditions {w.r.t. urban macro/micro cells} \cite{rappaport2017overview},\footnote{The border value $\gamma=2$ leads to $P_{\text{CELL}}^R=\frac{2 P^E}{(d_{\text{MAX}}^2-d_{\text{MIN}}^2) \cdot f^\eta \cdot c} \cdot \left[\ln(d_{\text{MAX}})-\ln(d_{\text{MIN}})\right]$, which  is however not further discussed in this work as $\gamma>2$ under commonly observed propagation conditions \cite{rappaport2017overview}.} thus obtaining the following closed-form \ac{RFP} expression:
\begin{eqnarray}
\label{eq:pavg_solved}
P_{\text{CELL}}^R = \frac{2 P^E}{(d_{\text{MAX}}^2-d_{\text{MIN}}^2) \cdot f^\eta \cdot c \cdot (\gamma - 2)} \cdot \left[\frac{1}{d_{\text{MIN}}^{(\gamma - 2)}} - \frac{1}{d_{\text{MAX}}^{(\gamma - 2)}}\right]
\end{eqnarray}

In this way, we have derived a simple model to compute the cell \ac{RFP} over the whole area served by a given \ac{gNB}. By observing in more detail Eq.~(\ref{eq:pavg_solved}), we can note that $P^R_{\text{CELL}}$ is strongly influenced by the propagation exponent $\gamma$ (as expected), which results into different scaling factors for the radiated power $P^E$. Other parameters that affect $P_{\text{CELL}}^R$ include the adopted frequency $f$, the constant term $c$, as well as the distances $d_{\text{MIN}}$ and $d_{\text{MAX}}$.




\subsection{RFP Model at Fixed Distance}

We then provide a second model to evaluate the \ac{RFP} from the serving \ac{gNB}. In particular, let us now consider a generic pixel at a fixed distance $d_{\text{FX}}$ from the serving \ac{gNB}. The \ac{RFP} $P^R_{\text{FX}}$ at distance $d_{\text{FX}}$ from the serving cell is then formally expressed as:

\begin{equation}
\label{eq:pfx}
P^R_{\text{FX}}=\frac{P^E}{d^{\gamma}_{\text{FX}}  \cdot f^\eta  \cdot c}.
\end{equation}

By varying $d_{\text{FX}}$ in the previous equation, we compute the \ac{RFP} over different evaluation points from the serving \ac{gNB}. For example, we can compute the \ac{RFP} in close proximity to the minimum distance $d_{\text{MIN}}$, whose pixels are subject to the highest \ac{RFP} levels.

\subsection{RFP Upper Bound from Neighboring Cells}
\label{sec:rfp_neigh_cells}

In the following, we provide a \ac{UB} to estimate the \ac{RFP} from $N^I$ neighboring \ac{gNB}. Let us denote with $d_{\text{SITE}}$ the inter-site distance between a pair of neighboring \acp{gNB}. By assuming that the sites hosting \acp{gNB} are positioned with a regular deployment, the relantionship between $d_{\text{SITE}}$ and $d_{\text{MAX}}$ is expressed as:
\begin{equation}
d_{\text{SITE}}=2 \zeta \cdot d_{\text{MAX}},
\end{equation} 
where $\zeta \in (0,1)$ is a geometric parameter that is introduced to avoid coverage holes over the whole territory under consideration. 
Our idea is then to assume a fixed contribution from each neighbor, which is computed at the edge of the serving \ac{gNB}. More formally, the \ac{RFP} from neighbors is expressed as:
\begin{equation}
\label{eq:neigh}
    P_{\text{NEIGH}}^R = N^I \frac{P^E}{ d_{\text{MAX}}^{\gamma}\cdot(2\zeta -1)^{\gamma} \cdot f^\eta \cdot c}.
\end{equation}

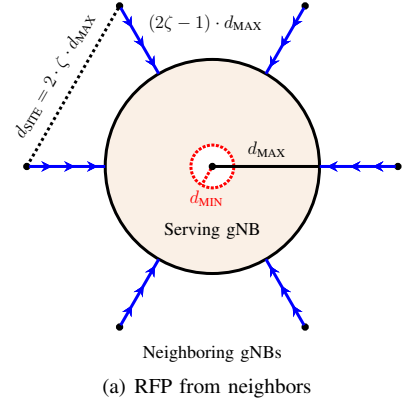
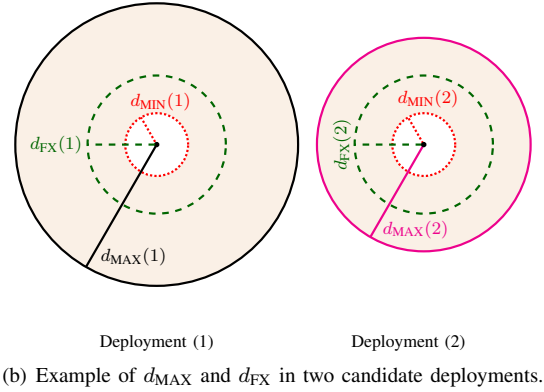
\begin{figure}[t]
\centering
\subfigure[\ac{RFP} from neighbors]
{
\resizebox{0.6\columnwidth}{!}{
  \begin{tikzpicture}[scale=1.0,>=latex]
	\begin{scope}
	[
		every node/.style={
		regular polygon, 
		regular polygon sides=6,
		draw,
		minimum width=0.95cm,
		line width=2pt,
		fill=white,
		outer sep=0,
		inner sep=0}
	]
	\definecolor{lightsalmon}{RGB}{222,222,222};
	\definecolor{gray}{RGB}{205,205,205};
        \definecolor{lightgray}{RGB}{240,240,240};
         \definecolor{linen}{RGB}{250,240,230};
  \definecolor{coral}{RGB}{255,127,80};
 
   \node[dummy] (dummy) {};
   


    \node[minimum width=8.66cm,draw=white] (bighex) {};
    \node[circle=black,draw, align=center, fill=black, minimum width=0.1cm] (bighexcenter) at (bighex)  {};
    \node[circle=black,draw, align=center, fill=black, minimum width=0.1cm] (center1) at (bighex.corner 1)  {};
    \node[circle=black,draw, align=center, fill=black, minimum width=0.1cm] (center2) at (bighex.corner 2)  {};
    \node[circle=black,draw, align=center, fill=black, minimum width=0.1cm] (center3) at (bighex.corner 3)  {};
    \node[circle=black,draw, align=center, fill=black, minimum width=0.1cm] (center4) at (bighex.corner 4)  {};
    \node[circle=black,draw, align=center, fill=black, minimum width=0.1cm] (center5) at (bighex.corner 5)  {};
    \node[circle=black,draw, align=center, fill=black, minimum width=0.1cm] (center6) at (bighex.corner 6)  {};

    \node[minimum width=5cm,draw=white, fill=none] (smallhex) at (bighex) {}; 
    \node[circle,minimum width=5cm,draw=black, fill=linen] (4avgtris) at (bighex) {}; 

    \node[minimum width=1cm,draw=white, fill=none] (dminhex) at (bighex) {}; 
    \node[circle,minimum width=1cm,draw=red,densely dotted, fill=white] (diminhexcircl) at (bighex) {}; 

    \draw[color=red, line width=2pt,densely dotted] (bighexcenter) --  node[draw=none,fill=none,color=red,yshift=-0.5cm,xshift=0cm] {\large $d_{\text{MIN}}$} (dminhex.corner 4); 

    \draw[color=blue, line width=2pt,arrow data={0.3}{stealth},arrow data={0.6}{stealth},arrow data={0.9}{stealth}] (center2) -- (4avgtris);
    \draw[color=blue, line width=2pt,arrow data={0.3}{stealth},arrow data={0.6}{stealth},arrow data={0.9}{stealth}] (center3) -- (4avgtris);
    \draw[color=blue, line width=2pt,arrow data={0.3}{stealth},arrow data={0.6}{stealth},arrow data={0.9}{stealth}] (center1) -- (4avgtris);
    \draw[color=blue, line width=2pt,arrow data={0.3}{stealth},arrow data={0.6}{stealth},arrow data={0.9}{stealth}] (center4) -- (4avgtris);
    \draw[color=blue, line width=2pt,arrow data={0.3}{stealth},arrow data={0.6}{stealth},arrow data={0.9}{stealth}] (center5) -- (4avgtris);
    \draw[color=blue, line width=2pt,arrow data={0.3}{stealth},arrow data={0.6}{stealth},arrow data={0.9}{stealth}] (center6) -- (4avgtris);

    \node[draw=none,fill=none,color=black,yshift=-0.7cm,xshift=2cm,yshift=0.3cm] at (center2) {\large $(2 \zeta - 1) \cdot d_{\text{MAX}}$};

    \draw[line width=2pt,dotted] (center2) -- node[rectangle,draw=none,fill=none,color=black,rotate=60,yshift=0.5cm] {\large $d_{\text{SITE}}=2 \cdot \zeta \cdot d_{\text{MAX}}$} (center3);

    \draw[color=black, line width=2pt] (bighexcenter) --  node[draw=none,fill=none,color=black,yshift=0.4cm,xshift=0cm] {\large $d_{\text{MAX}}$} (smallhex.corner 6); 

    \node[rectangle,draw=none,align=center,fill=none] (textserving) at ([yshift=-1.5cm]smallhex) {\large Serving \ac{gNB}};

    \node[rectangle,draw=none,align=center,fill=none] (textneighbors)  at ([yshift=-2.2cm]smallhex.south) {\large Neighboring \acp{gNB}};

    \node[circle,draw, align=center, fill=black, minimum width=0.1cm] (center) at (bighex) {};
    
  \end{scope}
	\end{tikzpicture}
}
	\label{fig:pollution_level}
}
\subfigure[Example of $d_{\text{MAX}}$ and $d_{\text{FX}}$ in two candidate deployments.]
{
\resizebox{0.8\columnwidth}{!}{
  \begin{tikzpicture}[scale=1.0,>=latex]
	\begin{scope}
	[
		every node/.style={
		regular polygon, 
		regular polygon sides=6,
		draw,
		minimum width=0.95cm,
		line width=1pt,
		outer sep=0,
		inner sep=0}
	]
	\definecolor{lightsalmon}{RGB}{222,222,222};
	\definecolor{gray}{RGB}{205,205,205};
        \definecolor{lightgray}{RGB}{240,240,240};
         \definecolor{linen}{RGB}{250,240,230};
  \definecolor{coral}{RGB}{255,127,80};
 
   \node[dummy] (dummy) {};
   


    \node[minimum width=4.5cm,draw=white] (bighex)  {};
    \node[circle,minimum width=4.5cm,draw=black, fill=linen] (bigcircle) at (bighex) {}; 

    \node[minimum width=1cm,draw=white, fill=none] (dminhex) at (bighex) {}; 
    \node[circle,minimum width=1cm,draw=red,densely dotted, fill=white] (diminhexcircl) at (bighex) {}; 
    \draw[color=red, line width=1pt,densely dotted] (bighexcenter) --  node[draw=none,fill=none,color=red,yshift=0.45cm,xshift=0.2cm] {\footnotesize $d_{\text{MIN}}(1)$} (dminhex.corner 2);

     

    
    \node[draw=none,minimum width=2.2cm, dashed] (dfx) {};

    \node[circle,minimum width=2.2cm,color=black!60!green, dashed] (dfxbis) {};
    
    
    \node[rectangle,draw=none,align=center,fill=none,text=black!60!green] (3text)  at ([yshift=0.0cm,xshift=-0.5cm]dfx.west) {\footnotesize  $d_{\text{FX}}(1)$};

    \node[circle,draw, align=center, fill=black, minimum width=0.05cm] (center) {};
    
    
    \draw[line width=1pt] (center) -- node[draw=none,fill=none,color=black,yshift=-0.8cm,xshift=0.2cm] {\footnotesize $d_{\text{MAX}}(1)$} (bighex.corner 4);
    
    \draw[line width=1pt, dashed,black!60!green] (center) -- (dfx.corner 3);

    \node[color=white,minimum width=3.4cm,draw=white] (smallhex) at ([xshift=2cm]bighex.east) {};

    \node[color=magenta, circle,minimum width=3.4cm,draw=magenta, fill=linen] (smallcircle) at (smallhex) {};

    \node[minimum width=1cm,draw=white, fill=none] (dminhex2) at (smallhex) {}; 
    \node[circle,draw, align=center, fill=black, minimum width=0.05cm] (center2) at (smallhex) {};

    \node[circle,minimum width=1cm,draw=red,densely dotted, fill=white] (diminhexcircl2) at (smallhex) {}; 

    \draw[color=red, line width=1pt,densely dotted] (center2) --  node[draw=none,fill=none,color=red,yshift=0.5cm,xshift=0.2cm] {\footnotesize $d_{\text{MIN}}(2)$} (dminhex2.corner 2);

    
    

     
     \node[draw=none,minimum width=2.2cm, dashed] (dfx2) at (smallhex)  {};

     \node[circle,minimum width=2.2cm, dashed,color=black!60!green] (dfx2bis) at (smallhex)  {};

     \draw[line width=1pt,draw=magenta] (center2) -- node[draw=none,fill=none,color=magenta,xshift=0.3cm,yshift=-0.6cm] {\footnotesize $d_{\text{MAX}}(2)$} (smallhex.corner 4);
     
    
    \node[rectangle,draw=none,align=center,fill=none,rotate=90,color=black!60!green] (6text)  at ([yshift=0.0cm,xshift=-0.2cm]dfx2.west) {\footnotesize  $d_{\text{FX}}(2)$};
     \draw[line width=1pt, dashed,black!60!green] (center2) -- (dfx2.corner 3);


    
    
    \node[rectangle,draw=none,align=center,fill=none] (7text)  at ([yshift=-2.2cm]dfx.south) {\footnotesize Deployment (1)};
    
    \node[rectangle,draw=none,align=center,fill=none] (8text)  at ([xshift=3.1cm]7text.east) {\footnotesize Deployment (2)};
    
         \node[circle,draw, align=center, fill=black, minimum width=0.05cm] (center3) at (smallhex) {};
  \end{scope}
	\end{tikzpicture}
}
	\label{fig:hex_example}
}
\caption{(Left) Graphical sketch of the distances appearing in the \ac{RFP} computation from the neighbors of Eq.~(\ref{eq:neigh}). The \ac{RFP} from neighboring \acp{gNB} is evaluated at $(2 \zeta -1) \cdot d_{\text{MAX}}$. (Right) $d_{\text{MAX}}$, $d_{\text{MIN}}$ and $d_{\text{FX}}$ in two candidate deployments. Clearly, $d_{\text{MAX}}(2)<d_{\text{MAX}}(1)$, while $d_{\text{FX}}(2)=d_{\text{FX}}(1)$ and $d_{\text{MIN}}(2)=d_{\text{MIN}}(1)$. Figures best viewed in colors.}
\label{fig:rfp_neigh_and_comp}
\end{figure}

To give more insight, Fig.~\ref{fig:pollution_level} reports a graphical example of the $d_{\text{MAX}}$ and $\zeta$ terms appearing  on the right-hand side of Eq.~(\ref{eq:neigh}), by assuming a hexagonal deployment of \ac{gNB} sites and $N^I=6$.  {Clearly, the value of $N^I$ depends on the chosen geometry for placing the \acp{gNB}, which in this case is an hexagonal grid. Therefore, Eq.~(\ref{eq:neigh}) includes the contributions of first-level neighbors. The contributions of $i$-th neighbors (where $i \geq 2$) is intentionally omitted, due to the following reasons: \textit{i}) the received power from neighbors is scaled by a factor proportional to $i^{\gamma - 1}$ (details are omitted due to the lack of space), thus making the contributions of neighbors mostly irrelevant starting from the second level onwards, and \textit{ii}) the vertical orientation of each cell employs electrical and/or mechanical tilting to concentrate the radiation over the coverage area and avoid unwanted interference towards the neighboring \acp{gNB}.}

By observing the figure, it is trivial to verify that Eq.~(\ref{eq:neigh}) represents a \ac{UB} of the \ac{RFP} for each neighbor $i \in \mathcal{I}^{\text{NEIGH}}$ and for each pixel $p$ falling inside the coverage area of the serving \ac{gNB}, because $(2 \zeta -1) \cdot d_{\text{MAX}} \leq d_{(p,i)}$. In addition, although this \ac{UB} may appear rather conservative at a first glance, we will show that the impact of neighbors on the total \ac{RFP} is always limited.


\subsection{RFP Ratio Among 5G Deployments}
In the final part of this section, we put together the previous models to evaluate the \ac{RFP} increase/decrease when comparing a pair of candidate 5G deployments.
Let us start by defining the total cell \ac{RFP} (including the \ac{UB} from neighbors) as:
\begin{equation}
P_{\text{TOT-CELL}}^R = P_{\text{CELL}}^R + P_{\text{NEIGH}}^R.
\end{equation}
In a similar way, the total \ac{RFP} at fixed distance is equal to:
\begin{equation}
\label{eq:rfp_fixed_distance}
P_{\text{TOT-FX}}^R = P_{\text{FX}}^R + P_{\text{NEIGH}}^R.
\end{equation}

Let us now consider two distinct 5G deployments, which are denoted by indexes $(1)$ and $(2)$, respectively. Each 5G deployment is characterized by specific settings in terms of densification level (and hence maximum coverage distance $d_{\text{MAX}}$), as well as the other parameters including e.g., the adopted frequency $f$, the propagation exponent $\gamma$, etc. We then define the cell \ac{RFP} ratio among the two deployments as:
\begin{equation}
\label{eq:ratio_cell}
\delta(P^R_{\text{TOT-CELL}})=\frac{P_{\text{TOT-CELL}}^R(1)}{P_{\text{TOT-CELL}}^R(2)}.
\end{equation}
In particular, when $\delta(P^R_{\text{TOT-CELL}})>1$, deployment (1) pollutes more than deployment (2). The opposite holds when $\delta(P^R_{\text{TOT-CELL}})<1$. On the other hand, when $\delta(P^R_{\text{TOT-CELL}})=1$, no change in the total cell \ac{RFP} is observed among the two deployments.  {Since an \ac{UB} is used to express the \ac{RFP} from neighbors, $\delta(P^R_{\text{TOT-CELL}})$ may naturally differ w.r.t. the ``real'' \ac{RFP} ratio (i.e., the one computed with the actual distance from neighbors, as in Eq.~(\ref{eq:gen_model})). To face this issue, we compare in this work the $\delta(P^R_{\text{TOT-CELL}})$ values computed from the model against the ``real'' \ac{RFP} ratio computed by simulation. The theoretical proof of the approximation introduced by Eq.~(\ref{eq:ratio_cell}) is left for future work. In general, the outcomes of our model may be useful to distinguish between the cases $\delta(P^R_{\text{TOT-CELL}})<1$ and $\delta(P^R_{\text{TOT-CELL}})>1$, while the results from simulation allows precisely quantifying the \ac{RFP} variation.} 

In a similar way, let us introduce the \ac{RFP} ratio at fixed distance as:
\begin{equation}
\label{eq:ratio_fixed_distance}
\delta(P^R_{\text{TOT-FX}})=\frac{P_{\text{TOT-FX}}^R(1)}{P_{\text{TOT-FX}}^R(2)}.
\end{equation}

By varying the parameters assigned to each deployment option, we are able to evaluate their effect on the \ac{RFP} ratios. For example, the impact of increasing the densification level is evaluated by imposing $d_{\text{MAX}}(2)<d_{\text{MAX}}(1)$. To this aim, a graphical sketch of two candidate deployments, subject to different densification levels, is shown in Fig.~\ref{fig:hex_example}. In this example, deployment (2) is denser than deployment (1), since $d_{\text{MAX}}(2)<d_{\text{MAX}}(1)$. However, $d_{\text{FX}}(2)=d_{\text{FX}}(1)$, i.e., the same observation point is assumed for the \ac{RFP} at fixed distance. In addition, $d_{\text{MIN}}(2)=d_{\text{MIN}}(1)$, i.e., the same exclusion zone is applied when computing the cell \ac{RFP}. In the following, we shed light on the adopted 5G scenarios.





\section{Description of 5G Scenarios}
\label{sec:scenario_definition}

\begin{table}[t]
	\caption{\ac{RFP} parameters setting over the different 5G scenarios.}
	\label{tab:scenarios}
	\scriptsize
\centering
\begin{tabular}{|>{\columncolor{Coral}}p{0.2cm}|p{0.83cm}|c|c|c|c|c|}
\hline
\rowcolor{Coral} & & \multicolumn{5}{c|}{\textbf{Scenario}}\\[-0.05em]
\rowcolor{Coral} \multirow{-2}{*}{\begin{sideways}\textbf{Set}\end{sideways}} & \multirow{-2}{*}{\begin{sideways}\textbf{Param.}\end{sideways}} & \textbf{S1} & \textbf{S2} & \textbf{S3} & \textbf{S4} & \textbf{S5}\\[-0.05em]
\hline
\rowcolor{Linen}\cellcolor{Coral} & $d_{\text{MIN}}$ & \multicolumn{5}{c|}{{5~[m]-}15~[m]}\\[-0.05em]
\cellcolor{Coral} & $d_{\text{MAX}}(1)$ & \multicolumn{5}{c|}{500~[m]}\\[-0.05em]
\rowcolor{Linen}\cellcolor{Coral} & $d_{\text{MAX}}(2)$ & 250~[m] & 100~[m] & 250~[m] & 500~[m] & 50~[m]\\[-0.05em]
\cellcolor{Coral} &  $\delta(d_{\text{MAX}})$ & 2 & 5 & 2 & 1 & 10\\[-0.05em]
\rowcolor{Linen}\cellcolor{Coral} &  $\gamma(1)$ & \multicolumn{5}{c|}{3}\\[-0.05em]
\multirow{-6}{*}{\begin{sideways}\textbf{Distance}\end{sideways}} \cellcolor{Coral} &  $\gamma(2)$ & 3 & 2.1 & \multicolumn{2}{c|}{3} & 2.1 \\[-0.05em]
\hline
\rowcolor{Linen}\cellcolor{Coral} &  $f(1)$ & \multicolumn{5}{c|}{0.7~[GHz]}\\[-0.05em]
\cellcolor{Coral} & $f(2)$ & \multicolumn{2}{c|}{0.7~[GHz]} & \multicolumn{3}{c|}{3.7~[GHz]}\\[-0.05em]
\rowcolor{Linen}\cellcolor{Coral} &  $\delta(f)$ & \multicolumn{2}{c|}{1} & \multicolumn{3}{c|}{0.19}\\[-0.05em]
\multirow{-4}{*}{\begin{sideways}\textbf{Frequency}\end{sideways}} \cellcolor{Coral} &  $\eta$ & \multicolumn{5}{c|}{2}\\[-0.05em]
\hline
\rowcolor{Linen}\cellcolor{Coral} &  $c(1)$ & \multicolumn{5}{c|}{32.4~[dB]}\\[-0.05em]
\cellcolor{Coral} &  $c(2)$ & \multicolumn{5}{c|}{32.4~[dB]}\\[-0.05em]
\rowcolor{Linen}\multirow{-3}{*}{\begin{sideways}\textbf{Baseline}\end{sideways}} \cellcolor{Coral} &  $\delta(c)$ & \multicolumn{5}{c|}{1}\\[-0.05em]
\hline
\cellcolor{Coral} & $\zeta$ & \multicolumn{5}{c|}{$\sqrt{3}/2$ (Hexagonal placement of \ac{gNB} sites)}  \\[-0.05em]
\rowcolor{Linen}\multirow{-2}{*}{\begin{sideways}\textbf{Neigh.}\end{sideways}} \cellcolor{Coral} & $N^I$ & \multicolumn{5}{c|}{$\{0,6\}$}  \\[-0.05em]
\hline
\cellcolor{Coral} & $P_{\text{TH}}^{R}(1)$ & \multicolumn{5}{c|}{-90~[dBm]}\\[-0.05em]
\rowcolor{Linen}\cellcolor{Coral} & $P_{\text{TH}}^{R}(2)$ & \multicolumn{3}{c|}{-90~[dBm]} & \multicolumn{2}{c|}{-87~[dBm]}\\[-0.05em]
\multirow{-3}{*}{\begin{sideways}\textbf{MSP}\end{sideways}} \cellcolor{Coral} & $\delta(P_{\text{TH}}^{R})$ & \multicolumn{3}{c|}{1} & \multicolumn{2}{c|}{0.5}\\[-0.05em]
\hline
\cellcolor{Coral} & $S_{\text{MAX}}$ & \multicolumn{5}{c|}{0.1~[W/m$^2$]}\\[-0.05em]
\rowcolor{Linen}\cellcolor{Coral} & $G_{\text{TX}}$ & \multicolumn{5}{c|}{15~[dB]}\\[-0.05em]
\multirow{-3}{*}{\begin{sideways}\textbf{ELP}\end{sideways}}  \cellcolor{Coral} & $L_{\text{TX}}$ & \multicolumn{5}{c|}{2.32~[dB]}\\
\hline
\end{tabular}
\vspace{-3mm}
\end{table}

Finding meaningful sets of input parameters to compare pairs of candidate 5G deployments is a fundamental step for the \ac{RFP} evaluation. Therefore, rather than considering a single scenario, which would narrow the scope of the presented outcomes, in this work we evaluate five representative 5G scenarios (denoted with S1-S5), detailed in Tab.~\ref{tab:scenarios}. In this way, we assess the impact of different densification levels (e.g, light, medium and strong) and other relevant parameters  {that are closely connected to densification, such as the} operating frequency $f$ and the propagation exponent $\gamma$. We then group the input parameters in Tab.~\ref{tab:scenarios} according to the following scopes: \textit{i}) distance, \textit{ii}) frequency, \textit{iii}) baseline path loss, \textit{iv}) neighbors, \textit{v}) \ac{MSP} setting, \textit{vi}) \ac{ELP} setting.

Let us first describe the main features of each parameter group. Focusing on the distance-related parameters, the $d_{\text{MAX}}$ values in the table are set {as follows: \textit{i}) $d_{\text{MAX}}=250$~[m] and lower values to represent the urban macro and dense urban deployments of 3GPP} \cite{3GPPscenarios}{, respectively; \textit{ii}) $d_{\text{MAX}}=500$~[m] to reflect sparser deployments (while still keeping $d_{\text{MAX}}$ lower than the distance triggering a change in the propagation exponent} \cite{rappaport2017overview}{).} Moreover, we introduce the parameter $\delta(d_{\text{MAX}})$ to denote the relative ratio between $d_{\text{MAX}}(1)$ and $d_{\text{MAX}}(2)$. The {values} of $d_{\text{MIN}}$ {are} set according to {typical sizes} of  \ac{gNB} exclusion {zones} (see e.g., \cite{colombi2020analysis} with theoretical maximum power). The {setting} of $d_{\text{MIN}}$ is also justified by assuming that each \ac{gNB} is hosted on the roof of a buildings and/or on top of a radio tower, and therefore the minimum distance between the \ac{UE} and the \ac{gNB} is not negligible.\footnote{{In this work, we always consider roof-mounted \acp{gNB}, for which the size of exclusion zone is not negligible. The evaluation of other types of \acp{gNB} without exclusion zones (e.g., indoor and/or femto \acp{gNB}) is left for future work.}} Eventually, the values of the propagation exponent $\gamma$ are set by assuming typical \ac{LOS}/\ac{NLOS} conditions. {In particular, the range $\gamma=(2,4]$ is adopted for the channel models of \ac{3GPP} in urban areas, i.e., 5G Urban Macro LOS/NLOS and 5G Urban Micro - Street Canyon} \cite{3GPPchannelmodels}. 
  
We then move our attention to the frequency-related parameters. We adopt the 5G Italian frequencies in the sub-6~[GHz] spectrum, which is the most promising option for offering coverage and a mixture of coverage and capacity.\footnote{The auction for 5G in Italy included also frequencies above 6~[GHz] (generally called ``mm-Waves''), which are however not treated in this work because the installation of \acp{gNB} operating on sub-6~[GHz] frequencies is prioritized w.r.t. equipment working on mm-Waves, due e.g., to 5G coverage constraints imposed by the national government. Therefore, it is expected that \acp{gNB} operating on mm-Waves will be not immediately and widely deployed. The evaluation of densification on mm-Waves is then left as a future work.}  Consequently, we set $f(1)=0.7$~[GHz] and $f(2)=\{0.7,3.7\}$~[GHz]. Moreover, we introduce the frequency ratio $\delta(f)=f(1)/f(2)$. Eventually, the $\eta$ exponent is set in accordance to \cite{rappaport2017overview}.

Focusing then on the parameters related to baseline path loss, we adopt $c=32.4$~[dB], in accordance with the \ac{FSPL} model reported by \cite{rappaport2017overview}. We remind that this term is derived from the constant part of the Friis' free space equation \cite{friis1946note}. Moreover, we introduce the ratio $\delta(c)=c(1)/c(2)$, which is equal to one in our scenarios.

We then analyze the parameters related to the \ac{RFP} from neighbors. We assume a hexagonal deployment of \acp{gNB} sites, thus resulting in $\zeta=\sqrt{3}/2$. In addition, we consider two distinct settings for the number of neighbors $N^I$ when computing the \ac{RFP}. More in depth, when $N^I=0$, the \ac{RFP} is solely due to the serving cell, while no contribution from the neighbors is assumed (thus representing an ideal case). On the other hand, when $N^I=6$, the \ac{RFP} includes the contributions from the six closest neighbors w.r.t. the serving \ac{gNB}.

We then detail the parameters to set the radiated power $P^E$. When an \ac{MSP} approach is assumed, we consider values of sensitivity threshold $P_{\text{TH}}^R$
 slightly higher than the minimum operating ones (defined in \cite{minsens} for the different 5G frequencies, bands and sub-carrier spacing). In addition, we consider cases in which $P_{\text{TH}}^R(2)>P_{\text{TH}}^R(1)$, i.e., deployment (2) enforces a better sensitivity, and hence a (potential) better service.  {More formally, the setting $P_{\text{TH}}^R=-90$~[dBm] allows guaranteeing the maximum throughput when up to 20~[MHz] of bandwidth (with 15~[kHz] of subcarrier spacing) is used in the 700~[MHz] frequency, and when up to 30 [Mhz] of bandwidth (with 15 [kHz] of subcarrier spacing) is used in the 3700~[MHz] frequency} \cite{minsens}{. In addition, the setting $P_{\text{TH}}^R=-87$~[dBm] allows achieving the maximum throughput up to 70 [MHz] of bandwidth (with 60 [kHz] of subcarrier spacing)} \cite{minsens}. Eventually, we introduce the ratio $\delta(P_{\text{TH}}^R)=P_{\text{TH}}^R(1)/P_{\text{TH}}^R(2)$, whose values are also reported in the table. On the other hand, when the radiated power is set according to \ac{ELP}, we assume the Italian \ac{PD} limit for residential areas, i.e., $S_{\text{MAX}}=0.1$~[W/m$^2$], which we remind is much more stringent that the ones defined in international guidelines, such as the \ac{ICNIRP} 2020 limits \cite{ICNIRPGuidelines:20}. Moreover, the values of $G_{\text{TX}}$ and $L_{\text{TX}}$ are set in accordance to \cite{itutk70}.

In the following step, we provide a comparative description among the two candidate deployments in S1-S5:
\begin{itemize}
\item[S1)] Light densification scenario. The only parameter (slightly) changing across deployment (1) and deployment (2) is $d_{\text{MAX}}$ (and consequently $\delta(d_{\text{MAX}})$). In S1, deployment (2) is slightly denser than deployment (1), while all the other parameters do not vary across the two deployments;
\item[S2)]  Moderate densification scenario, which is subject to a radical variation of $d_{\text{MAX}}$ and $\gamma$ across the two deployments. In S2, the operator adopts a denser deployment in (2) compared to (1). This choice is coupled with a different site deployment strategy and/or site configuration setting, which allows a better coverage over the territory. Consequently, $\gamma(2)<\gamma(1)$;
\item[S3)]  Light densification with frequency change. In S3, both $d_{\text{MAX}}$ and $f$ are varied in the two deployments. Specifically, while the 0.7~[GHz] frequency in (1) is primarily used to provide coverage, the 3.7~[GHz] frequency of deployment (2) allows achieving a good mixture of coverage and capacity. Moreover, we consider a slight reduction of  $d_{\text{MAX}}(2)$ compared to $d_{\text{MAX}}(1)$;
\item[S4)] No densification with frequency change. In S4, $d_{\text{MAX}}$ is not varied, while $f$ is increased when passing from deployment (1) to deployment (2). When an \ac{MSP} setting is assumed, this scenario also imposes $P_{\text{TH}}^R(2)>P_{\text{TH}}^R(1)$. With these settings, the operator is able to support a 5G service demanding a higher amount of capacity in deployment (2) compared to (1);
\item[S5)] Strong densification with frequency change. In S5 we impose $d_{\text{MAX}}(2)\ll d_{\text{MAX}}(1)$, $f(2)>f(1)$, and $\gamma(2)<\gamma(1)$. As a consequence, the impact of passing from a sparse set of 5G \acp{gNB} to a very dense deployment is evaluated. Clearly, this choice has an impact on the propagation conditions, as users in deployment (2) tend to be in \ac{LOS} conditions w.r.t. the serving 5G \ac{gNB}, thus resulting in $\gamma(2)<\gamma(1)$. Moreover, when $P^E$ is set in accordance with \ac{MSP}, this scenario imposes an increase of the minimum sensitivity $P_{\text{TH}}^R$ in deployment (2) compared to (1). Similarly to S4, in S5 the operator provides a larger capacity to the users.
\end{itemize}

\begin{table*}[t]
\caption{Closed-form expressions for \ac{RFP} ratio at fixed distance $\delta (P^R_{\text{TOT-FX}})$ and cell \ac{RFP} ratio $\delta (P^R_{\text{TOT-CELL}})$ in the different scenarios when $N^I=0$ (MSP case). Table best viewed in colors.}
\label{tab:received_power_ratio_min_sens}
\scriptsize
\centering
\renewcommand\arraystretch{1.2} 
\begin{tabular}{|c|c|c|c|c|}
\hline
\rowcolor{Linen} & \multicolumn{4}{c|}{\textbf{RFP Ratio}}\\
\cline{2-5}
\rowcolor{Linen} & \multicolumn{2}{c|}{\textbf{at Fixed Distance $\delta (P^R_{\text{TOT-FX}})$}}  & \multicolumn{2}{c|}{\textbf{Cell $\delta (P^R_{\text{TOT-CELL}})$}}\\ 
\cline{2-5}
\rowcolor{Linen} & & \textbf{\ac{RFP}} & & \textbf{\ac{RFP}} \\ 
\rowcolor{Linen} \multirow{-4}{*}{\begin{sideways}\textbf{Scenario}\end{sideways}}& \multirow{-2}{*}{\textbf{Formula}} & \textbf{Increase?} & \multirow{-2}{*}{\textbf{Formula}} & \textbf{Increase?} \\ 
\hline
 \cellcolor{Linen} &\cellcolor{Green} &\cellcolor{Green} & & \\[-1.4em]
 \cellcolor{Linen} \textbf{S1} &\cellcolor{Green} $\delta(d_{\text{MAX}})^3$ &\cellcolor{Green} No  & $ \delta(A)^{-1}  \cdot \delta(d_{\text{MAX}})^2 \cdot \left[ \frac{d_{\text{MAX}}(1) -d_{\text{MIN}} }{d_{\text{MAX}}(2) -d_{\text{MIN}} }\right]$ & Num. Eval.\\[0.5em]
\hline
\cellcolor{Linen} &\cellcolor{Green} &\cellcolor{Green} & & \\[-1.4em]
 \cellcolor{Linen} \textbf{S2} &\cellcolor{Green} $\frac{\delta(d_{\text{MAX}})^{\gamma(2)}}{\beta(1)^{\gamma(1)-\gamma(2)}}$ &\cellcolor{Green} No  & $\delta(A)^{-1}  \cdot \frac{d_{\text{MAX}}(1)^{\gamma(1)}}{d_{\text{MAX}}(2)^{\gamma(2)}} \cdot \frac{\gamma(2)-2}{\gamma(1)-2} \cdot \left[\frac{d_{\text{MIN}}^{(2-\gamma(1))}-d_{\text{MAX}}(1)^{(2-\gamma(1))}}{d_{\text{MIN}}^{(2-\gamma(2))}-d_{\text{MAX}}(2)^{(2-\gamma(2))}}\right]$ & Num. Eval. \\[1.2em] 
\hline
\cellcolor{Linen} &\cellcolor{Green} &\cellcolor{Green} & & \\[-1.4em]
 \cellcolor{Linen} \textbf{S3} &\cellcolor{Green} $\delta(d_{\text{MAX}})^{3}$ &\cellcolor{Green} No & $ \delta(A)^{-1}  \cdot \delta(d_{\text{MAX}})^2 \cdot \left[ \frac{d_{\text{MAX}}(1) -d_{\text{MIN}} }{d_{\text{MAX}}(2) -d_{\text{MIN}} }\right]$ & Num. Eval.\\[0.5em] 
\hline
\rowcolor{Red} \cellcolor{Linen} & & & & \\[-1.4em]
\rowcolor{Red} \cellcolor{Linen} \textbf{S4} & $ \delta({P^R_{\text{TH}}})$ & Yes & $ \delta(P_{\text{TH}}^R)$ & Yes \\[0.5em] 
\hline
 \cellcolor{Linen} & & & & \\[-1.4em]
 \cellcolor{Linen} \textbf{S5} & $\frac{\delta(d_{\text{MAX}})^{\gamma(2)}\cdot \delta({P^R_{\text{TH}}})}{\beta(1)^{\gamma(1)-\gamma(2)}}$ & Num. Eval. & $\delta(P_{\text{TH}}^R)\cdot \delta(A)^{-1}  \cdot \frac{d_{\text{MAX}}(1)^{\gamma(1)}}{d_{\text{MAX}}(2)^{\gamma(2)}} \cdot \frac{\gamma(2)-2}{\gamma(1)-2} \cdot \left[\frac{d_{\text{MIN}}^{(2-\gamma(1))}-d_{\text{MAX}}(1)^{(2-\gamma(1))}}{d_{\text{MIN}}^{(2-\gamma(2))}-d_{\text{MAX}}(2)^{(2-\gamma(2))}}\right]$ & Num. Eval. \\[1.0em] 
\hline
\end{tabular}
\vspace{-3mm}
\end{table*}

\section{RFP Evaluation}
\label{sec:evaluation}

We initially focus on the closed-form expressions for the \ac{RFP} to scientifically analyze the impact of densification{, by considering the same size $d_{\text{MIN}}$ of the exclusion zone  across the candidate deployments}. We then move our attention to the numerical evaluation of the \ac{RFP} in order to give more insights about the impact of the input parameters on the obtained \ac{RFP} values. {In the following step, we investigate the impact of varying $d_{\text{MIN}}$ across the deployments. Eventually, we shed light on the impact of alternative policies (based on the spectrum allocation) to set the radiated power. Finally, we evaluate the \ac{RFP} when the number of neighbors is increased}.

\subsection{Closed-Form RFP Evaluation}

As a first step, we retrieve the closed-form expressions for the \ac{RFP} by assuming $N^I=0$ and the \ac{MSP} setting. We remind that, in this way, each \ac{gNB} solely pollutes its own coverage area, i.e., no \ac{RFP} from neighbors is assumed. In addition, the emitted power is set in order to guarantee the minimum sensitivity thresholds $P^R_{\text{TH}}(1)$ and $P^R_{\text{TH}}(2)$. Let us then introduce the ratio among the coverage areas of a single \ac{gNB} in the two deployments as:
\begin{equation}
 \delta( A ) = \frac{\pi(d_{\text{MAX}}(1)^2 -d_{\text{MIN}}^2)}{\pi(d_{\text{MAX}}(2)^2 -d_{\text{MIN}}^2)}. 
\end{equation}
In addition, let us introduce the $\beta$ parameter, which is defined as the ratio between the observation point at fixed distance and the maximum coverage distance. More formally, we have
$\beta(1)=\frac{d_{\text{FX}}(1)}{d_{\text{MAX}}(1)}$ for deployment (1). Unless otherwise specified, we set: {$d_{\text{MIN}}=15$~[m]}  {for both the deployments (i.e., the same exclusion zone is assumed)}, {$d_{\text{FX}}(1)=d_{\text{MIN}}+1$ [m]}, {$d_{\text{FX}}(2)=d_{\text{FX}}(1)$}, i.e., the \ac{RFP} at fixed distance is evaluated in close proximity to $d_{\text{MIN}}$.

Tab.~\ref{tab:received_power_ratio_min_sens} reports the \ac{RFP} ratio at fixed distance $\delta(P^R_{\text{TOT-FX}})$ and the cell \ac{RFP} ratio $\delta(P^R_{\text{TOT-CELL}})$ over the different scenarios. The color of each cell in the table is set according to the following rule: \textit{i}) green if the expression leads to an \ac{RFP} decrease, \textit{ii}) red if the expression is lower than one, and hence an \ac{RFP} increase is experienced, \textit{iii}) white if the \ac{RFP} expression includes terms that are respectively lower and higher than one, thus requiring a numerical evaluation.

By analyzing the colors of the table, we can note that the \ac{RFP} at fixed distance is decreased in S1-S3, as the mathematical expressions for $\delta(P^R_{\text{TOT-FX}})$ include terms that are all greater than unity (i.e., $\delta(d_{\text{MAX}})^{\gamma(2)}$, $1/\beta(1)^{\gamma(2)-\gamma(1)}$). This is a first important outcome, which proves that  densification does not always trigger an increase in the \ac{RFP}, in contrast to the population's belief. On the other hand, the \ac{RFP} at fixed distance is increased in S4, i.e., $\delta(P^R_{\text{TOT-FX}})<1$. Nevertheless, the \ac{RFP} increase in this scenario depends solely on the sensitivity thresholds ratio $\delta(P_{\text{TH}}^R)$, and hence it can be easily controlled by the operator. In addition, we remind that S4 imposes both frequency and minimum sensitivity increases in deployment (2) w.r.t. deployment (1), while the level of densification is not varied. Eventually, the \ac{RFP} at fixed distance in S5 includes terms that are greater than unity (i.e., $1/\beta(1)^{\gamma(2)-\gamma(1)}$ and $\delta(d_{\text{MAX}})^{\gamma(2)}$) and other ones that are instead lower than unity (i.e., $\delta(P_{\text{TH}}^R)$). Therefore, an approach based on numerical evaluation is required, in order to assess the impact on the \ac{RFP}.

Focusing then on the expressions for the cell \ac{RFP} ratio (right part of Tab.~\ref{tab:received_power_ratio_min_sens}), we can note that the terms $\delta(A)^{-1}$ and $\delta(d_{\text{MAX}})=\frac{d_{\text{MAX}}(1)}{d_{\text{MAX}}(2)}$  appear in S1, S2, S3, and S5. Since $\delta(A)^{-1}$ is lower than unity while $\delta(d_{\text{MAX}})$ is greater than one, a numerical evaluation is required in order to assess the overall impact on the \ac{RFP} ratio. On the other hand, $\delta(P^R_{\text{TOT-CELL}})=\delta(P_{\text{TH}}^R)$ in S4, and consequently the same considerations already reported for $\delta(P^R_{\text{TOT-FX}})$ hold also in this case. 

\begin{table*}[t]
\caption{Closed-form expressions for \ac{RFP} ratio at fixed distance $\delta (P^R_{\text{TOT-FX}})$ and cell \ac{RFP} ratio $\delta (P^R_{\text{TOT-CELL}})$ in the different scenarios when $N^I=0$ (ELP case). Table best viewed in colors.}
\label{tab:received_power_ratio_constant}
\scriptsize
\centering
\renewcommand\arraystretch{1.2} 
\begin{tabular}{|c|c|c|c|c|}
\hline
\rowcolor{Linen} & \multicolumn{4}{c|}{\textbf{RFP Ratio}}\\
\cline{2-5}
\rowcolor{Linen} & \multicolumn{2}{c|}{\textbf{at Fixed Distance $\delta (P^R_{\text{TOT-FX}})$}}  & \multicolumn{2}{c|}{\textbf{Cell $\delta (P^R_{\text{TOT-CELL}})$}}\\ 
\cline{2-5}
\rowcolor{Linen} & & \textbf{\ac{RFP}} & & \textbf{\ac{RFP}} \\ 
\rowcolor{Linen} \multirow{-4}{*}{\begin{sideways}\textbf{Scenario}\end{sideways}}& \multirow{-2}{*}{\textbf{Formula}} & \textbf{Increase?} & \multirow{-2}{*}{\textbf{Formula}} & \textbf{Increase?} \\ 
\hline
 \cellcolor{Linen} &\cellcolor{Green} &\cellcolor{Green} & & \\[-1.4em]
 \cellcolor{Linen} \textbf{S1} &\cellcolor{Green} 1 & \cellcolor{Green}No & $ \delta( A )^{-1} \cdot \delta(d_{\text{MAX}})^{-1} \cdot \frac{d_{\text{MAX}}(1)-d_{\text{MIN}}}{d_{\text{MAX}}(2)-d_{\text{MIN}}}$ & Num. Eval. \\[0.5em]
\hline
 \cellcolor{Linen} &\cellcolor{Red} &\cellcolor{Red} & & \\[-1.4em]
 \cellcolor{Linen} \textbf{S2} &\cellcolor{Red} $(\beta(1)\cdot d_{\text{\text{MAX}}}(1))^{\gamma(2)-\gamma(1)}$ &\cellcolor{Red}Yes & $\delta(A)^{-1} \cdot \frac{\gamma(2)-2}{\gamma(1)-2} \cdot \left[\frac{d_{\text{MIN}}^{(2-\gamma(1))}-d_{\text{MAX}}(1)^{(2-\gamma(1))}}{d_{\text{MIN}}^{(2-\gamma(2))}-d_{\text{MAX}}(2)^{(2-\gamma(2))}}\right]$ &Num. Eval. \\[1.2em] 
\hline
\cellcolor{Linen} &\cellcolor{Green} &\cellcolor{Green} & & \\[-1.4em]
\cellcolor{Linen} \textbf{S3} &\cellcolor{Green} $\delta(f)^{-\eta}$ &\cellcolor{Green}No &$ \delta( A )^{-1} \cdot \delta(f)^{-\eta} \cdot \delta(d_{\text{MAX}})^{-1} \cdot \frac{d_{\text{MAX}}(1)-d_{\text{MIN}}}{d_{\text{MAX}}(2)-d_{\text{MIN}}}$ & Num. Eval. \\[0.5em] 
\hline
\rowcolor{Green}\cellcolor{Linen} & & & & \\[-1.4em]
\rowcolor{Green}\cellcolor{Linen} \textbf{S4} & $\delta(f)^{-\eta}$ & No & $ \delta(f)^{-\eta}  $  & No \\[0.5em] 
\hline
 \cellcolor{Linen} & & & & \\[-1.4em]
 \cellcolor{Linen} \textbf{S5} & $ (\beta(1)\cdot d_{\text{\text{MAX}}}(1))^{\gamma(2)-\gamma(1)} \cdot \delta(f)^{-\eta}$ & Num. Eval. & $\delta(f)^{-\eta} \cdot \delta(A)^{-1} \cdot \frac{\gamma(2)-2}{\gamma(1)-2} \cdot \left[\frac{d_{\text{MIN}}^{(2-\gamma(1))}-d_{\text{MAX}}(1)^{(2-\gamma(1))}}{d_{\text{MIN}}^{(2-\gamma(2))}-d_{\text{MAX}}(2)^{(2-\gamma(2))}}\right]$ & Num. Eval. \\[1.0em] 
\hline
\end{tabular}
\vspace{-4mm}
\end{table*}

In the following, we analyze the impact of the \ac{ELP} setting on the \ac{RFP} levels. We remind that \ac{ELP} adjusts $P^E$ in order to ensure the maximum \ac{PD} limit at the border of the exclusion zone (i.e., at $d_{\text{MIN}}$), and therefore the radiated power does not scale with the maximum coverage distance $d_{\text{MAX}}$. Tab.~\ref{tab:received_power_ratio_constant} reports the closed-form expressions for $\delta (P^R_{\text{TOT-FX}})$ and $\delta (P^R_{\text{TOT-CELL}})$, by assuming $N^I=0$. Interestingly, $\delta (P^R_{\text{TOT-FX}})$ is equal to one in S1, meaning that the \ac{RFP} {at fixed distance} is unchanged when passing from deployment (1) to deployment (2). On the other hand, the \ac{RFP} tends to decrease in S3 and S4, since the term $\delta(f)^{-\eta}$ is clearly higher than unity. However, there are also cases in which the \ac{RFP} ratio at fixed distance is increased in deployment (2) w.r.t. deployment (1). For example, $\delta (P^R_{\text{TOT-FX}})<1$ in S2, since the term $(\beta(1)\cdot d_{\text{MAX}})^{\gamma(2)-\gamma(1)}$ is lower than unity. In addition, a numerical evaluation of the formula is required in S5, since $(\beta(1)\cdot d_{\text{MAX}})^{\gamma(2)-\gamma(1)}<1$ and $\delta(f)^{-\eta}>1$. Focusing then on the cell \ac{RFP}, $\delta (P^R_{\text{TOT-CELL}})$ has to be numerically evaluated in S1, S2, S3, and S5 to assess the \ac{RFP} increase/decrease. However, the cell \ac{RFP} is surely decreased in S4, as the \ac{RFP} ratio is equal to $\delta(f)^{-\eta}$, which is greater than one. 

We then move our attention to the closed-form expressions of \ac{RFP} when the contributions from neighbors are considered, i.e., $N^I>0$. Intuitively, when introducing the $P^{R}_{\text{NEIGH}}$ term of Eq.~(\ref{eq:neigh}) in the \ac{RFP} ratios, the closed-form expressions of $\delta (P^R_{\text{TOT-FX}})$ and $\delta (P^R_{\text{TOT-CELL}})$ become more complex than the $N^I=0$ case reported in Tab.~\ref{tab:received_power_ratio_min_sens}-\ref{tab:received_power_ratio_constant}. As a result, it is not possible to grasp the \ac{RFP} increase/decrease by simply analyzing the terms appearing in the closed-form expressions. Therefore, rather than reporting such {(complex)} expressions, we directly compute $\delta (P^R_{\text{TOT-FX}})$ and $\delta (P^R_{\text{TOT-CELL}})$ from our models by applying the input parameters and then we analyze the outcomes in the following subsection.

\subsection{Numerical Evaluation of the \ac{RFP}}

We move our attention to the numerical evaluation of the \ac{RFP} obtained by our models. This step is mandatory for the following reasons. First, it is possible to grasp the actual values of \ac{RFP} increase/decrease, including the cases in Tab.~\ref{tab:received_power_ratio_min_sens}-\ref{tab:received_power_ratio_constant} for which the \ac{RFP} variation could not be preliminary determined. Second, we thoroughly evaluate the case with $N^I>0$, i.e., the \ac{RFP} includes the contributions from neighbors.

More formally, we adopt Eq.~(\ref{eq:pavg_solved}),(\ref{eq:pfx}),(\ref{eq:neigh})-(\ref{eq:ratio_fixed_distance}) to compute $\delta (P^R_{\text{TOT-FX}})$ and $\delta (P^R_{\text{TOT-CELL}})$ across the different scenarios, by selectively imposing $N^I=\{0,6\}$ and $P^E$ according to \ac{MSP} or \ac{ELP} setting. In order to better position the outcomes derived by our models, we also build a simple simulator coded in Matlab R2020a software that allows us to compute the \ac{RFP} $P^R_{(p)}$ for each pixel $p$ of the coverage area, by applying Eq.~(\ref{eq:gen_model}). To this aim, we assume a tessellation of non-overlapping squared pixels, each of them with $1\times1$~[m$^2$] size. In addition, the pixel to \ac{gNB} distance is computed from the center of the pixel. Focusing then on the computation of the \ac{RFP} from neighboring \acp{gNB} done in the simulator, we assume the following cases: \textit{i}) $N^I=0$, in which $P^R_{(p)}$ is solely computed from the serving \ac{gNB}, \textit{ii}) $N^I=6$, in which $P^R_{(p)}$ includes the \ac{RFP} from the serving \ac{gNB} \textit{and} the \ac{RFP} from the six closest neighboring \acp{gNB} w.r.t. the serving one. Given the \ac{RFP} $P^R_{(p)}$ in each pixel, we then extract: \textit{i}) the average \ac{RFP} at fixed distance $d_{\text{FX}}$, computed over the pixels that are at distance $d_{(p,s)}$ within $d_{\text{FX}} - \epsilon \leq d_{(p,s)} \leq d_{\text{FX}} + \epsilon$, with $\epsilon=1$~[m]; \textit{ii}) the average cell \ac{RFP}, computed over the pixels at distance $d_{\text{MIN}} \leq d_{(p,s)}  \leq d_{\text{MAX}}$; \textit{iii}) the \ac{RFP} ratio at fixed distance and the cell \ac{RFP} ratio, given the averages in \textit{i}) and \textit{ii}), computed over deployment (1) and deployment (2) for scenarios S1-S5.

\begin{figure*}[t]
\centering
\subfigure[$\delta(P^R_{\text{TOT-FX}})$, MSP setting]
{
	\includegraphics[width=7cm]{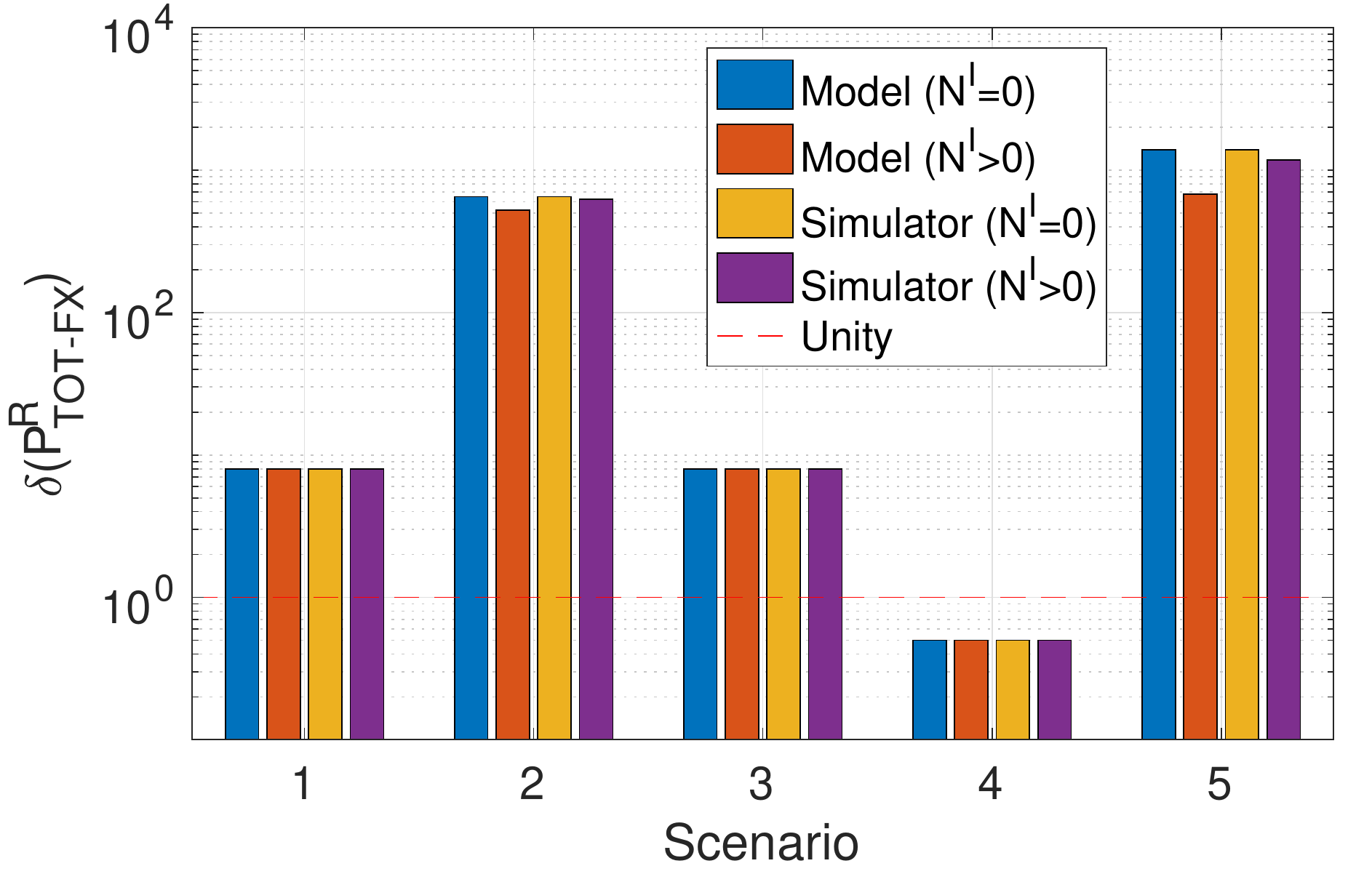}
	\label{fig:delta_p_r_tot_fix_pe_min_sens}
}
\subfigure[$\delta(P^R_{\text{TOT-FX}})$, ELP setting]
{
	\includegraphics[width=7cm]{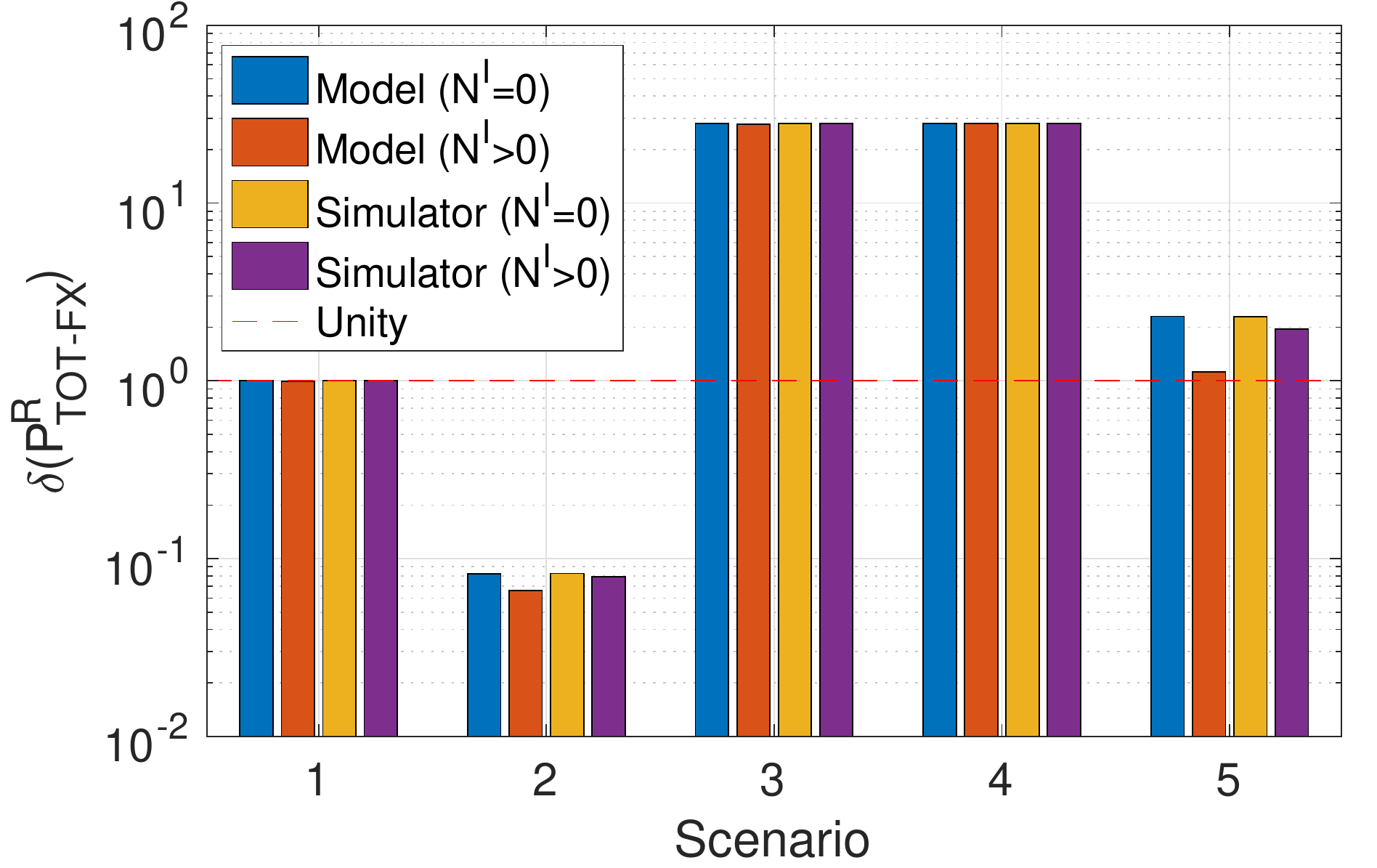}
	\label{fig:delta_p_r_tot_fix_pe_constant}
}

\centering
\subfigure[$\delta(P^R_{\text{TOT-CELL}})$, MSP setting]
{
	\includegraphics[width=7cm]{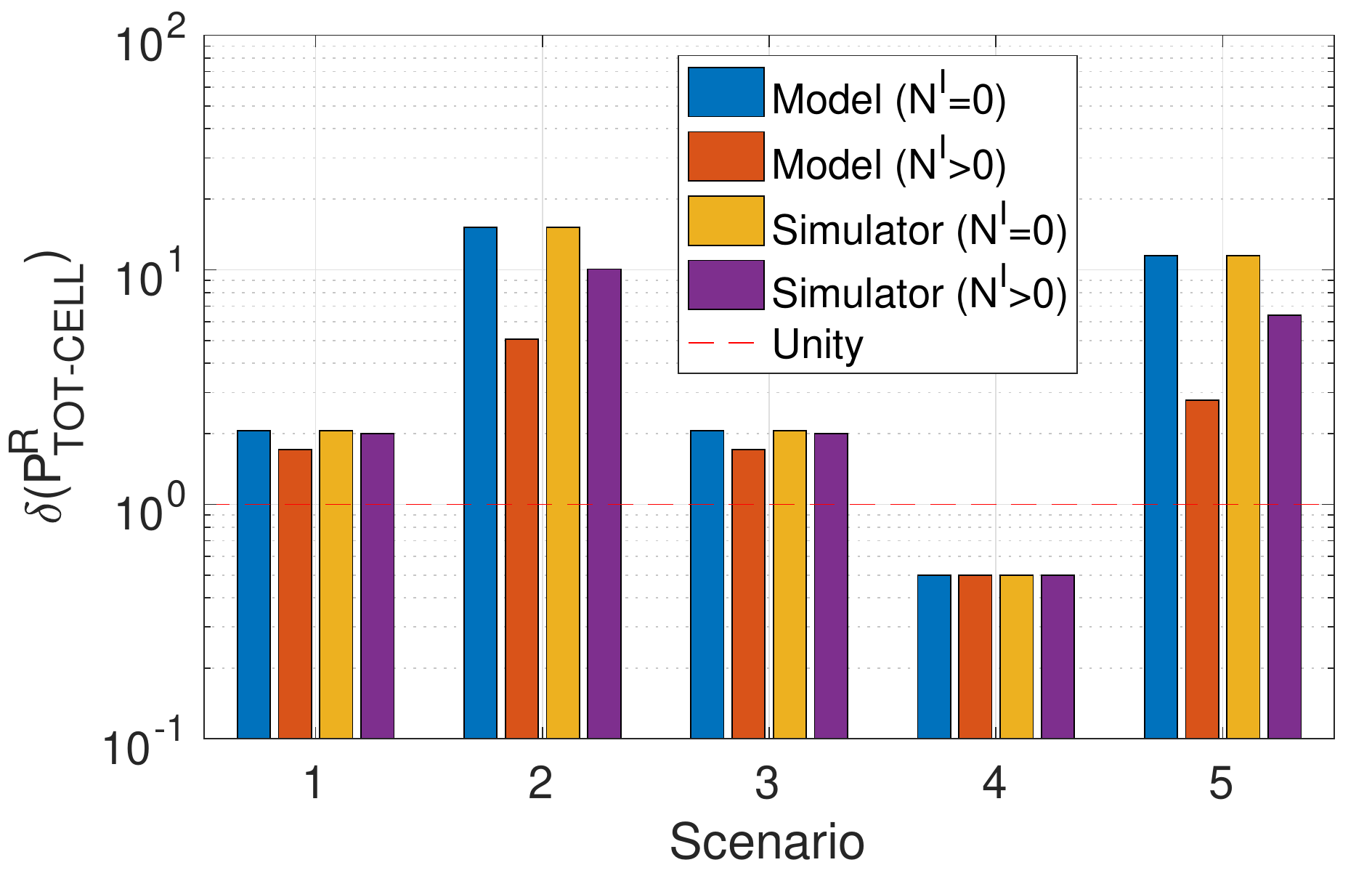}
	\label{fig:delta_p_r_tot_cell_pe_min_sens}
}
\subfigure[$\delta(P^R_{\text{TOT-CELL}})$, ELP setting]
{
	\includegraphics[width=7cm]{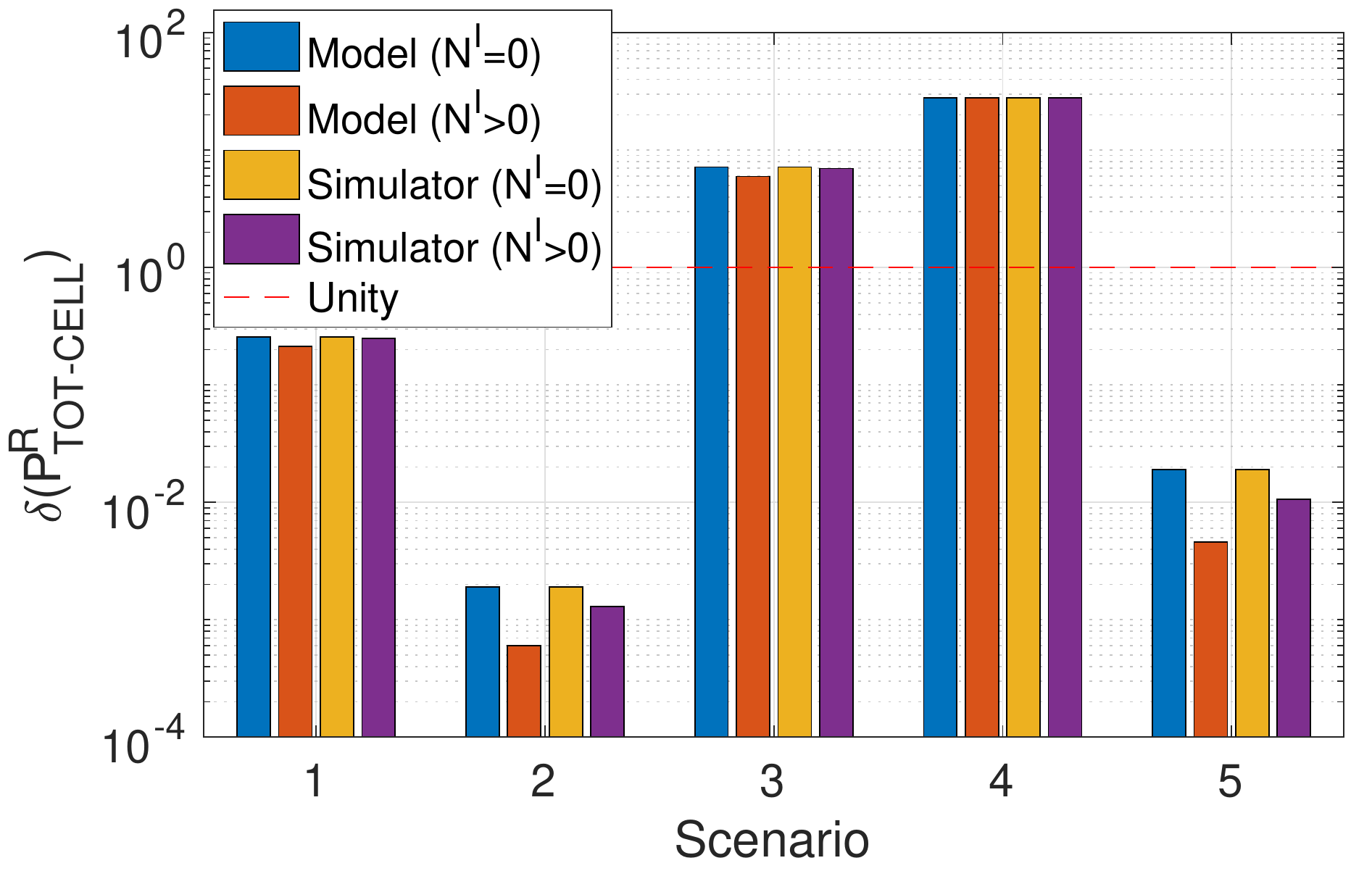}
	\label{fig:delta_p_r_tot_cell_pe_constant}
}
\caption{Numerical evaluation of \ac{RFP} by considering: \ac{RFP} ratio at fixed distance $\delta(P^R_{\text{TOT-FX}})$ vs. cell \ac{RFP} ratio $\delta(P^R_{\text{TOT-CELL}})$, \ac{ELP} vs. \ac{MSP} power setting, number of neighbors $N^I$ equal to 0 or 6, model vs. simulation outcomes, and scenarios S1-S5. Figures best viewed in colors.}
\label{fig:num_eval}
\end{figure*}

Fig.~\ref{fig:num_eval} reports the numerical evaluation of the \ac{RFP} ratios, by considering the \ac{MSP} and \ac{ELP} settings, the impact of neighbors, and the model vs. simulator outcomes. Before going into the details of each scenario, let us remind that the results obtained with our models for the $N^I>0$ cases represent a worst-case, since the \ac{RFP} from neighbors is conservatively evaluated with the \ac{UB} of Eq.~(\ref{eq:neigh}). We now analyze $\delta(P^R_{\text{TOT-FX}})$ for the \ac{MSP} case, shown in Fig.~\ref{fig:delta_p_r_tot_fix_pe_min_sens}. Obviously, the \ac{RFP} ratio {at fixed distance} is larger than unity for S1, S2, S3 (in accordance with Tab.~\ref{tab:received_power_ratio_min_sens}). Astonishingly, the \ac{RFP} decrease is huge for all the scenarios introducing densification, since $\delta(P^R_{\text{TOT-FX}})\gg1$. We remind that, in this case, we evaluate the \ac{RFP} variation at distance $d_{\text{FX}}$ from the serving \ac{gNB}, i.e., close to $d_{\text{MIN}}$ in our setting. Eventually, scenario S4 confirms the previously reported outcomes, i.e., a controlled \ac{RFP} increase that depends on $\delta(P^R_{\text{TH}})$. Finally, the numerical evaluation over S5 reveals a strong \ac{RFP} reduction at fixed distance, i.e., {up to around three orders} of magnitude. Therefore, we can state that a strong densification dramatically reduces the \ac{RFP} when this metric is evaluated in proximity to $d_{\text{MIN}}$.

We then move our attention to the \ac{RFP} ratio at fixed distance for the \ac{ELP} setting, visualized in Fig.~\ref{fig:delta_p_r_tot_fix_pe_constant}. The results over scenarios S1, S3, S4 confirm the findings reported in Tab.~\ref{tab:received_power_ratio_constant}, with a huge \ac{RFP} decrease in S3 and S4 (i.e., more than one order of magnitude in deployment (2) w.r.t. (1)). On the other hand, the \ac{RFP} is increased in S2 (as expected). Finally, the numerical evaluation demonstrates that the \ac{RFP} is reduced also in S5. By globally analyzing the outcomes of Fig.~\ref{fig:delta_p_r_tot_fix_pe_constant}, we can state that densification with \ac{ELP} setting does not increase the \ac{RFP} at fixed distance in all scenarios except from S2. 

In the following step, we evaluate $\delta(P^R_{\text{TOT-CELL}})$, reported in Fig.~\ref{fig:delta_p_r_tot_cell_pe_min_sens} and in Fig.~\ref{fig:delta_p_r_tot_cell_pe_constant} for the \ac{MSP} and \ac{ELP} settings, respectively. Focusing on the \ac{MSP} case (Fig.~\ref{fig:delta_p_r_tot_cell_pe_min_sens}), we can note that densification always reduces the average cell \ac{RFP}, since $\delta(P^R_{\text{TOT-CELL}})>1$ in S1, S2, S3, S5. Again, this is an important outcome that contradicts the common belief of the population about exponential increase of \ac{RFP}. In particular, the \ac{RFP} decrease triggered by densification can be huge, i.e., {around} one order of magnitude in deployment (2) w.r.t. deployment (1) {(scenarios S2 and S5)}. Clearly, $\delta(P^R_{\text{TOT-CELL}})<1$ in S4, in accordance with Tab.~\ref{tab:received_power_ratio_constant}. Focusing then on the \ac{ELP} case (Fig.~\ref{fig:delta_p_r_tot_cell_pe_constant}), the cell \ac{RFP} is decreased in S3 and S4. Not surprisingly, densification tends to increase the cell \ac{RFP} in scenarios S1, S2 and S5. With the \ac{ELP} setting, in fact, $P^E$ does not scale with $d_{\text{MAX}}$, thus resulting in $\delta(P^R_{\text{TOT-CELL}})>1$ when densification is applied (except from scenario S3, which couples a light densification to a frequency increase).

Eventually, we compare the outcomes from our models (bars with label ``model'' in Fig.~\ref{fig:num_eval})  w.r.t. the ones from the simulation (bars with label ``simulation'' in Fig.~\ref{fig:num_eval}). Interestingly, our models for the \ac{RFP} ratio nicely match the outcomes from simulations when $N^I=0$. On the other hand, when $N^I>0$, the \ac{RFP} ratios predicted by the models are in general lower than the ones computed through simulation. This is however an expected result, because the former {adopts} the \ac{UB} at distance $(2\zeta-1)\cdot d_{\text{MAX}}$, while the latter {exactly computes the \ac{RFP} for each distance} $d_{(p,i)}$.


\begin{figure}[t]
\centering
\subfigure[S5 - MSP setting]
{
	\includegraphics[width=7cm]{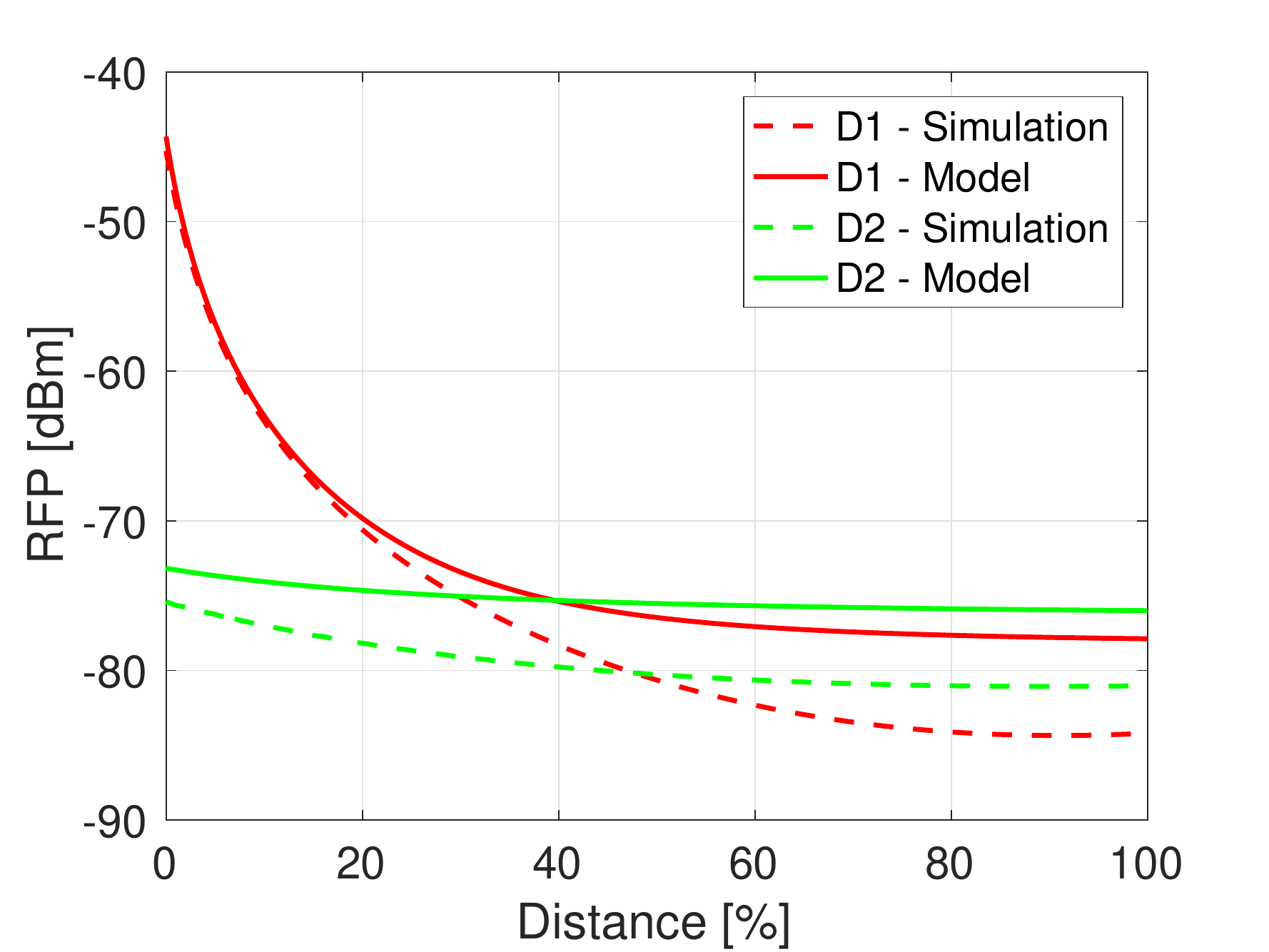}
	\label{fig:model_sim_comp_s5_MSP}
}
\subfigure[S3 - ELP setting]
{
	\includegraphics[width=7cm]{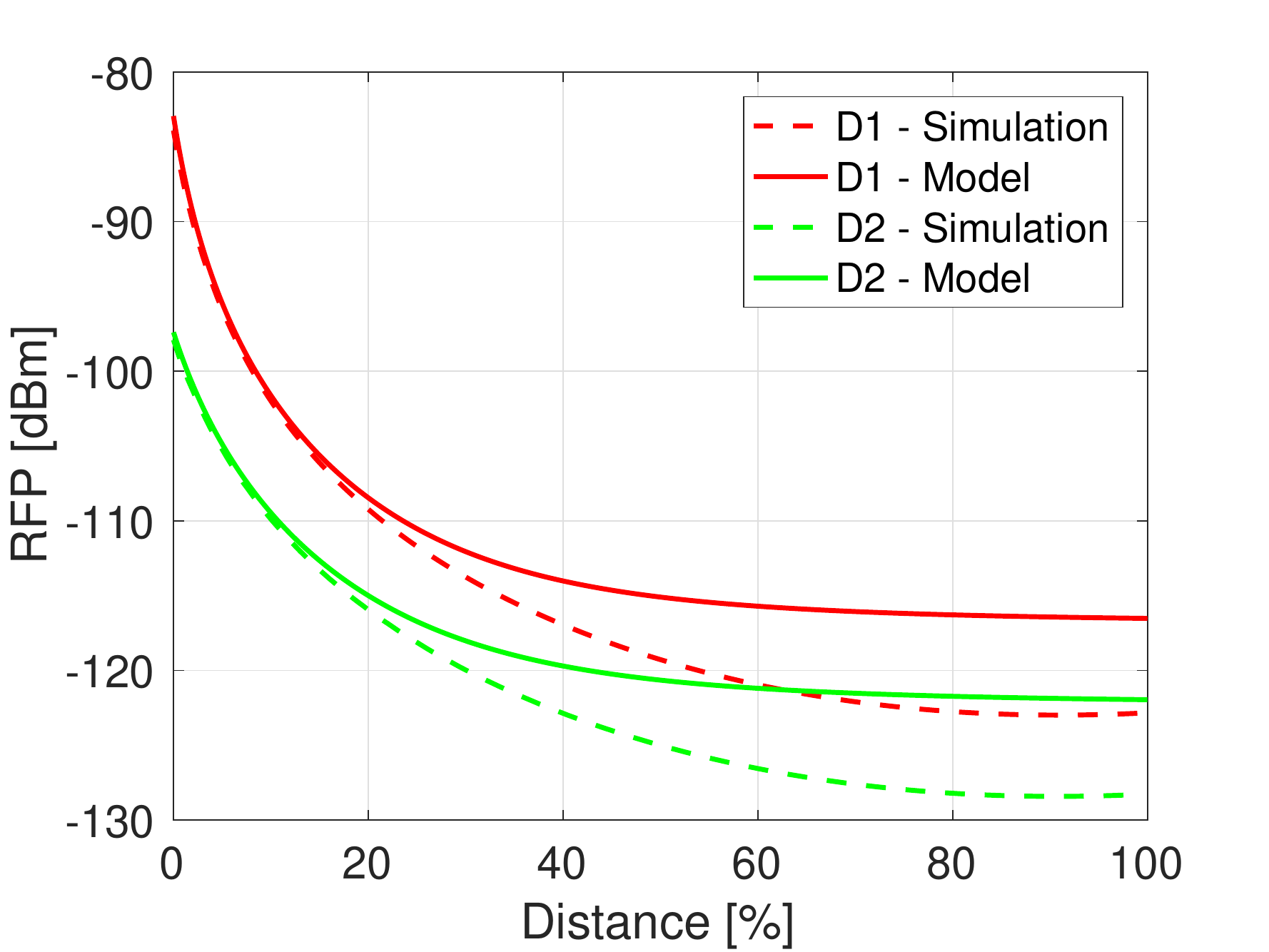}
	\label{fig:model_sim_comp_s3_ELP}
}
\caption{\ac{RFP} at fixed distance $P^R_{\text{TOT-FX}}$ in deployment (1) and deployment (2), computed with our model and by simulation. The figure reports $P^R_{\text{TOT-FX}}$ vs. the distance percentage for S5 with \ac{MSP} setting (left) and for S3 with \ac{ELP} (right).}
\label{fig:model_sim_comp}
\vspace{-5mm}
\end{figure}

We then analyze in depth the differences between the \ac{RFP} obtained from our models and the one computed from simulations. More in detail, we consider the variation of the fixed distance $d_{\text{FX}}$ and its impact on the \ac{RFP} model of Eq.~(\ref{eq:rfp_fixed_distance}). To this aim, we vary $d_{\text{FX}}$ between $d_{\text{MIN}}$ and $d_{\text{MAX}}$. Fig.~\ref{fig:model_sim_comp} reports the \ac{RFP} at fixed distance  $P^R_{\text{TOT-FX}}$ in deployment (1) and in deployment (2) for scenario S5 with \ac{MSP} setting (Fig.~\ref{fig:model_sim_comp_s5_MSP}) and S2 with \ac{ELP} setting (Fig.~\ref{fig:model_sim_comp_s3_ELP}). The outcomes of our model are derived by imposing $N^I=6$ neighboring \acp{gNB}. The figures report on the x-axis the distance percentage at which the \ac{RFP} is evaluated, being 0\% corresponding to $d_{\text{FX}}=d_{\text{MIN}}$ and 100\% equal to the maximum one, i.e., $d_{\text{FX}}=d_{\text{MAX}}$. In addition to the outcomes from our models, we include the \ac{RFP} from simulations, which is computed in this way: \textit{i}) we generate a set of bins, each of them is size 1~[m], between $d_{\text{MIN}}$ and $d_{\text{MAX}}$, \textit{ii}) we compute the \ac{RFP} for each pixel as per Eq.~(\ref{eq:gen_model}), \textit{iii}) we assign each pixel to the bin that includes the pixel distance $d_{(p,s)}$, \textit{iv}) we compute the average \ac{RFP} for each bin, and finally \textit{v}) we include in the plots of Fig.~\ref{fig:model_sim_comp} only the bins having non-zero \ac{RFP} values.

Several considerations hold by analyzing Fig.~\ref{fig:model_sim_comp}. First of all, the {values of} $P^R_{\text{TOT-FX}}$ {obtained from our model are} always higher than the {ones computed} from simulation. This result is expected, due to the \ac{UB} from neighbors. In particular, the difference (in [dBm]) between the model and the simulation is proportional to the distance percentage, with lower \ac{RFP} evaluation distances translating into smaller relative differences among model and simulation results and even overlapping \ac{RFP} values. This is also an expected outcome, since we remind that, in proximity to the serving \ac{gNB}, the \ac{RFP} contributions from neighbors are in general negligible w.r.t. the \ac{RFP} of the serving \ac{gNB}. On the other hand, there are cases in which the \ac{RFP} from neighbors is over-estimated by our model over the whole cell extent (e.g., deployment (2) in Fig.~\ref{fig:model_sim_comp_s5_MSP}). This is especially true for scenarios, like S5, in which the coverage size is extremely narrowed. In this way, we corroborate the findings of our models, which are derived under conservative and worst-case assumptions. Eventually, we can observe that $P^R_{\text{TOT-FX}}$ is a monotonic decreasing function w.r.t. the distance for both the model and the simulation (obviously). Finally, the comparison among deployment (1) and deployment (2)  reveals that the \ac{RFP} of the former is even lower than the latter for $\frac{d_{\text{FX}}}{d_{\text{MAX}}}\geq 0.4$ in S5 with \ac{MSP} setting (Fig.~\ref{fig:model_sim_comp_s5_MSP}). On the other hand, for lower distances, the absolute \ac{RFP} values are dramatically higher in deployment (1) w.r.t. (2). In addition, the slope of the \ac{RFP} is notably increased as $d_{\text{FX}}$ approaches $d_{\text{MIN}}$ in deployment (1). Such details trigger another important consideration: densification is extremely effective in reducing the \ac{RFP} for the users living in proximity to the installed \acp{gNB} \textit{and} to achieve a uniform \ac{RFP} distribution over the territory.

\begin{figure}[t]
\centering
\subfigure[S5 - Deployment (1)]
{
	\includegraphics[width=7cm]{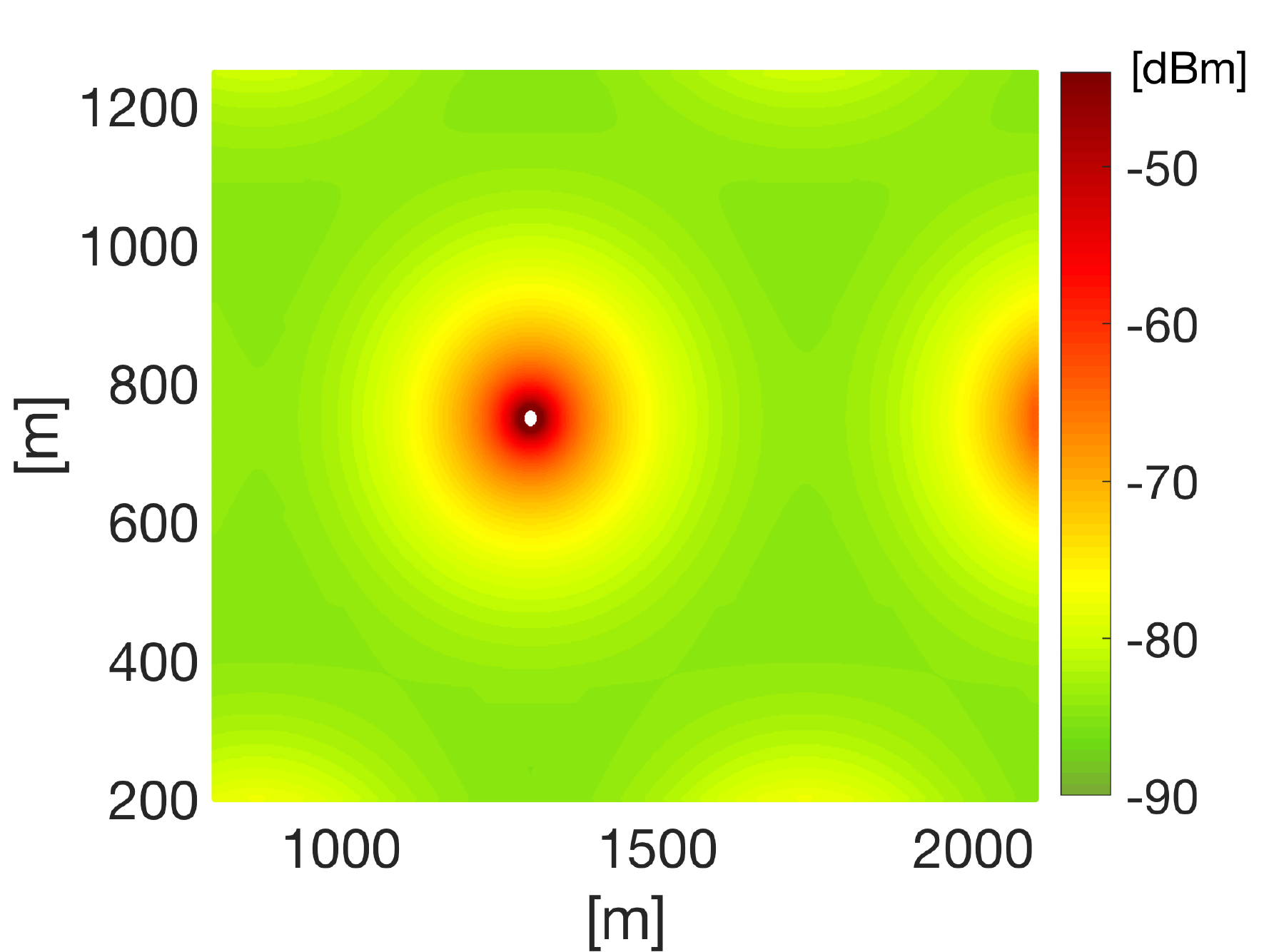}
	\label{fig:s5-1_res}
}
\subfigure[S5 - Deployment (2)]
{
	\includegraphics[width=7cm]{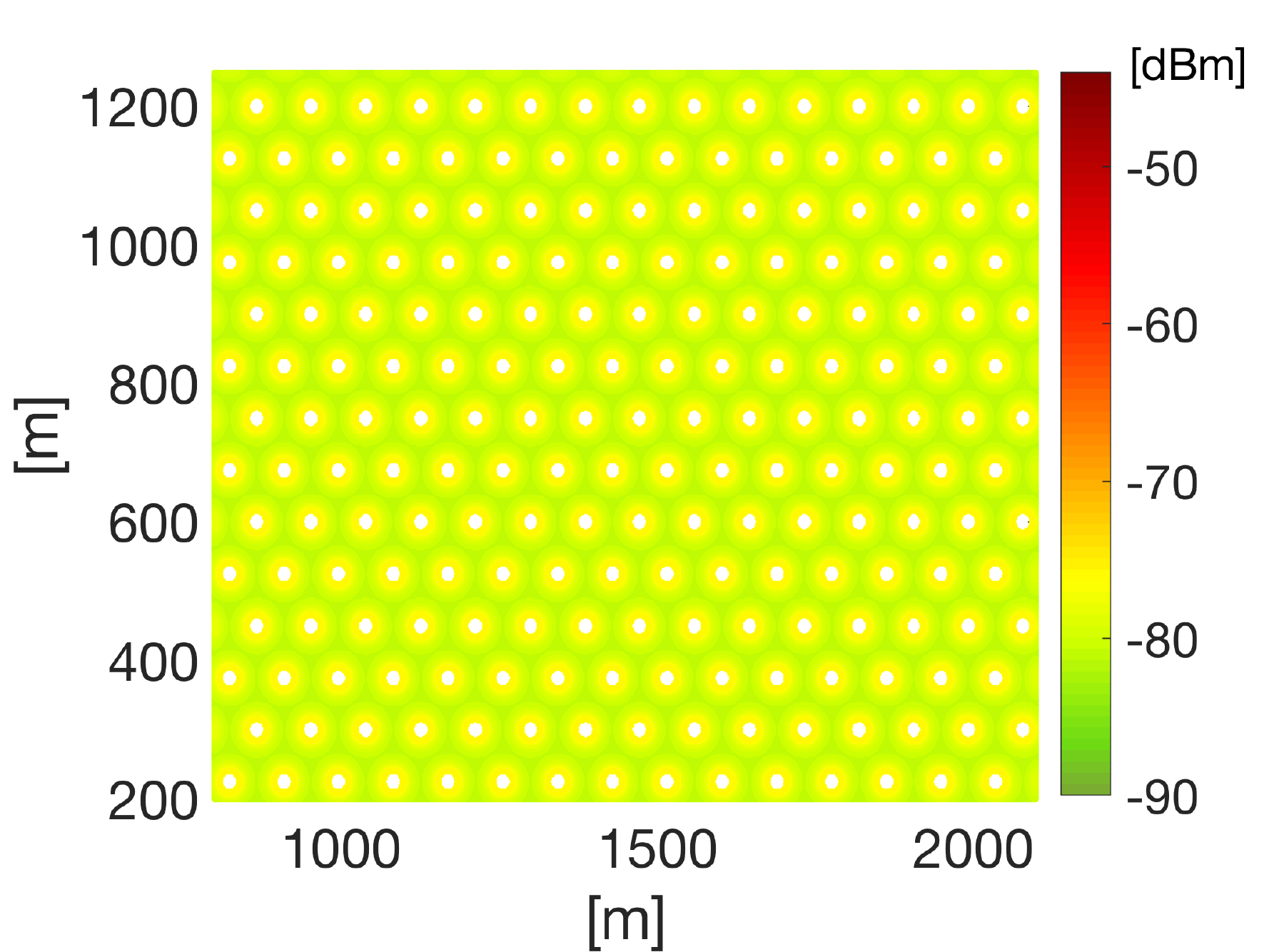}
	\label{fig:s5-2_res}
}
\caption{\ac{RFP} (in [dBm]) in scenario S5 for deployment 1 (left) and deployment 2 (right).}
\label{fig:s1-s5-hex_res_2}
\end{figure}

{In the following step, we provide a visual representation of} the pixel \ac{RFP} $P^R_{(p)}$, computed by simulation from Eq.~(\ref{eq:gen_model}) in scenario S5 with $N^I=6$ and \ac{MSP} setting. Fig.~\ref{fig:s1-s5-hex_res_2} reports the obtained results over the two deployments. By comparing this figure against the popular belief of the population that was sketched in Fig.~\ref{eq:gen_model}, several considerations can be drawn. First of all, the deployment of few \acp{gNB} does not necessarily mean low \ac{RFP} over the territory, as the \ac{RFP} tends to be pretty {high in} the zones that are close to the installed \acp{gNB}. For example, Fig.~\ref{fig:s5-1_res} shows that deployment (1) introduces huge \ac{RFP} levels in proximity to the installed \acp{gNB} (i.e., the orange and the red zones in the figure), because the radiated power $P^E$ is tuned to ensure the minimum sensitivity threshold $P^{R}_{\text{TH}}(1)$ at the cell edge $d_{\text{MAX}}(1)$. In addition, deploying a dense set of \acp{gNB} is not translated into an uncontrolled and exponential increase of \ac{RFP}. In Fig.~\ref{fig:s5-2_res}, in fact, the orange and red zones with large \ac{RFP} completely disappear, since the coverage size of each cell is shrank in deployment (2) w.r.t. deployment (1), and hence $P^E$ is now set to guarantee the minimum sensitivity at $d_{\text{MAX}}(2)<d_{\text{MAX}}(1)$. In addition, we remind that deployment (2) in this case also includes an increase of $P^{R}_{\text{TH}}$, and hence a better 5G service w.r.t. deployment (1). Despite this fact, however, the \ac{RFP} appears more uniform in deployment (2) w.r.t. deployment (1).

\subsection{{Exclusion zone variation}}

{When evaluating the \ac{RFP} in scenario S5, a natural observation is that the exclusion zone for deployment (2) is pretty large compared to the coverage size. Although the rationale for choosing the size of the exclusion zone is clear (e.g., we adopt the same equipment type and a rooftop installation over the two candidate deployments), an interesting step would be to assess the impact on the \ac{RFP} when the size of the exclusion zone in deployment (2) is reduced. More formally, let us denote the minimum distance with $d_{\text{MIN}}(1)$ and $d_{\text{MIN}}(2)$, respectively for deployment (1) and deployment (2). In addition, let us denote with $\delta(d_{\text{MIN}})$ the ratio among $d_{\text{MIN}}(1)$ and $d_{\text{MIN}}(2)$. Since $d_{\text{MIN}}(1) \neq d_{\text{MIN}}(2)$, we compare the two candidate deployments over two distinct fixed distances $d_{\text{FX}}(1)$ and $d_{\text{FX}}(2)$ (e.g., close to $d_{\text{MIN}}(1)$ for deployment (1) and close to $d_{\text{MIN}}(2)$ for deployment (2)).}

\begin{figure}[t]
\centering
\subfigure[$\delta(P^R_{\text{TOT-FX}})$]
{
	\includegraphics[width=7cm]{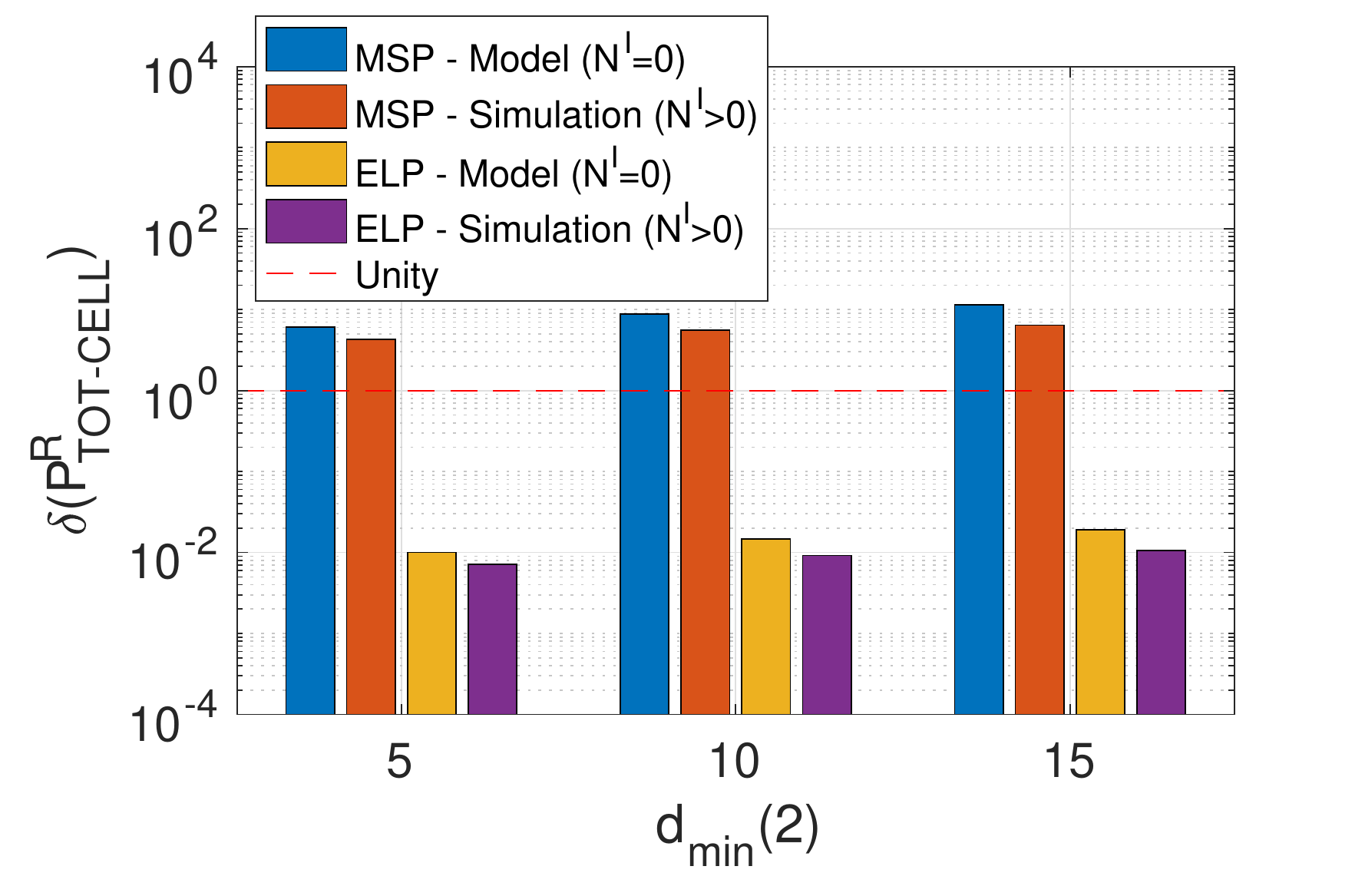}
	\label{fig:d_min_fix}
}
\subfigure[$\delta(P^R_{\text{TOT-CELL}})$]
{
	\includegraphics[width=7cm]{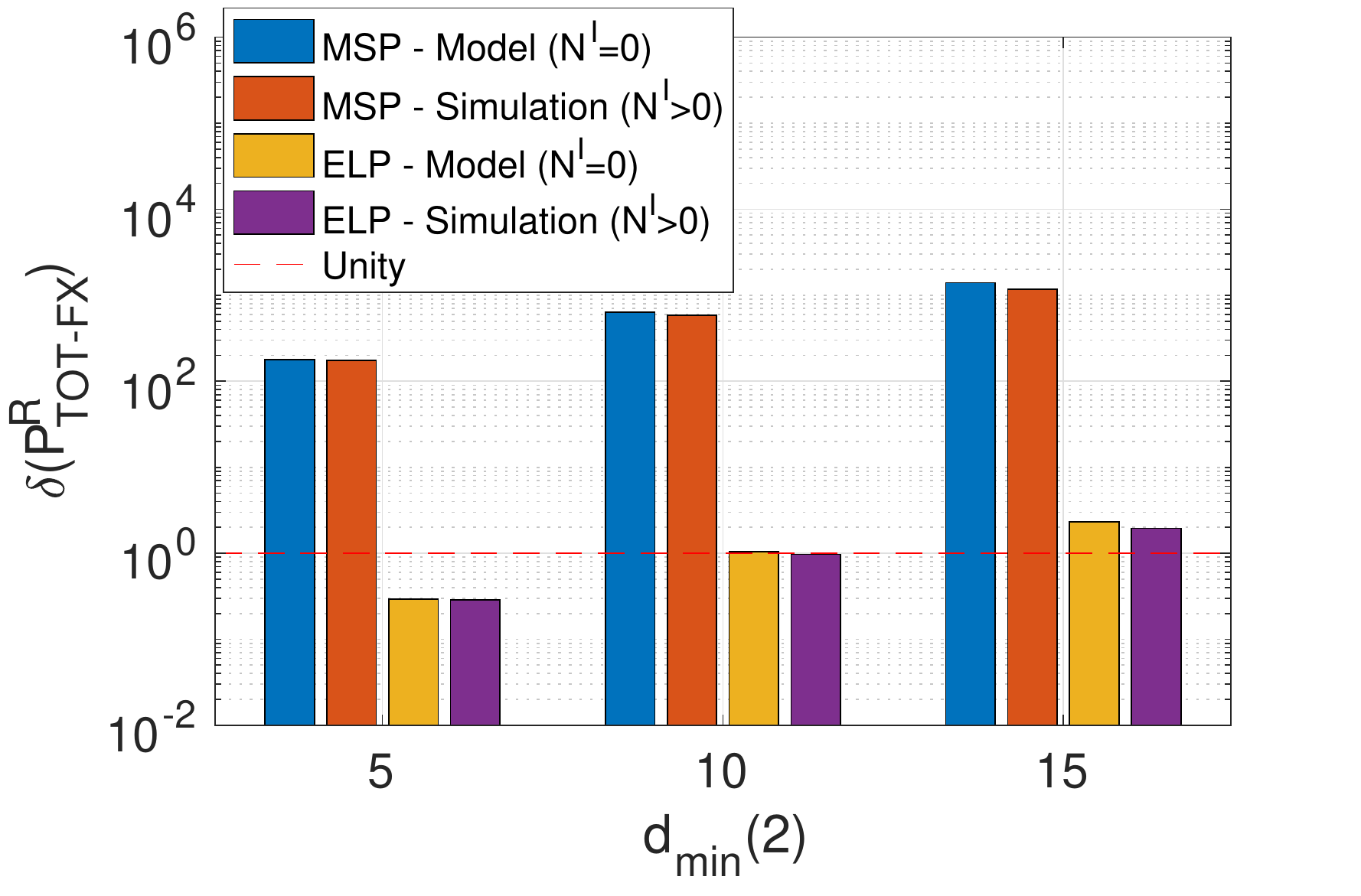}
	\label{fig:d_min_avg}
}
\caption{{Impact of $d_{\text{MIN}}(2)$ variation on $\delta(P^R_{\text{TOT-FX}})$ and $\delta(P^R_{\text{TOT-CELL}})$ (S5 scenario).}}
\label{fig:d_min}
\vspace{-4mm}
\end{figure}

{Fig.~\ref{fig:d_min} reports the numerical evaluation of \ac{MSP} and \ac{ELP} policies vs. different values of $d_{\text{MIN}}(2)$ (where $d_{\text{FX}}(2)$=$d_{\text{MIN}}(2)+1$~[m]), while all the other parameters are kept unchanged w.r.t. the already presented ones. Interestingly, the decrease of $d_{\text{MIN}}(2)$ tends to reduce both $\delta(P^R_{\text{TOT-FX}})$ and $\delta(P^R_{\text{TOT-CELL}})$. This effect is expected for \ac{MSP}, since the effective area covered by each cell in deployment (2) is increased as $d_{\text{MIN}}(2)$ is decreased. A similar trend is also observed in \ac{ELP}. In particular, $\delta(P^R_{\text{TOT-FX}})$ passes from values larger than unity when $d_{\text{MIN}}(2)=15$~[m] to values lower than unity $d_{\text{MIN}}(2)=5$~[m]. As a consequence, densification is not always beneficial in reducing the level of pollution at fixed distance in this case.}
 
\subsection{{Impact of spectrum-based power setting policies}}

{A natural question is the following: Which is the impact of adopting other policies in setting the \ac{gNB} radiated power? To answer such question, we have introduced a} \ac{SPS}{, based on the idea that the maximum radiated power is proportional to the amount of spectrum managed by the \ac{gNB}. Actually, \ac{SPS}-based policies are defined by different regulation authorities (e.g., } \ac{FCC}{) to  set the maximum power radiated by the \ac{gNB}, which is used during the planning of the network, e.g., when requesting permissions to install the \ac{gNB}. For example, \ac{FCC} defines a standardization 47~[dBm]/10~[MHz] for outdoor base stations that have to be installed at a height above 6~[m] from ground level} \cite{FCCstd} {However, we point out that the actual power radiated by the \ac{gNB} in operation is typically order of magnitudes lower than the maximum one} \cite{colombi2020analysis}{, thus resulting in a potential large over-estimation of \ac{RFP}. Therefore, in the following, we evaluate the impact of \ac{SPS}, by keeping in mind that the \ac{RFP} may be largely over-estimated in this case. More formally, the power radiated by each \ac{gNB} is then set to $P^E = P^F \cdot B/10$, where $P^F$ is the maximum amount of radiated power over 10~[MHz] of bandwdith, and $B$ is the adopted bandwidth by the deployment (in MHz). We then express the RFP ratio at fixed distance $\delta(P^R_{\text{TOT-FX}})$ and cell RFP ratio $\delta(P^R_{\text{TOT-CELL}})$ by considering the \ac{SPS} setting. Clearly, it holds that $P^E(1)=P^E(2)$ when $f(1)=f(2)$ (since $B(1)=B(2)$). Otherwise, when $f(1)\neq(2)$, $B(1)\neq B(2)$ and consequently $P^E(1)\neq P^E(2)$.}

\begin{figure}[t]
\centering
\subfigure[$\delta(P^R_{\text{TOT-FX}})$]
{
	\includegraphics[width=7cm]{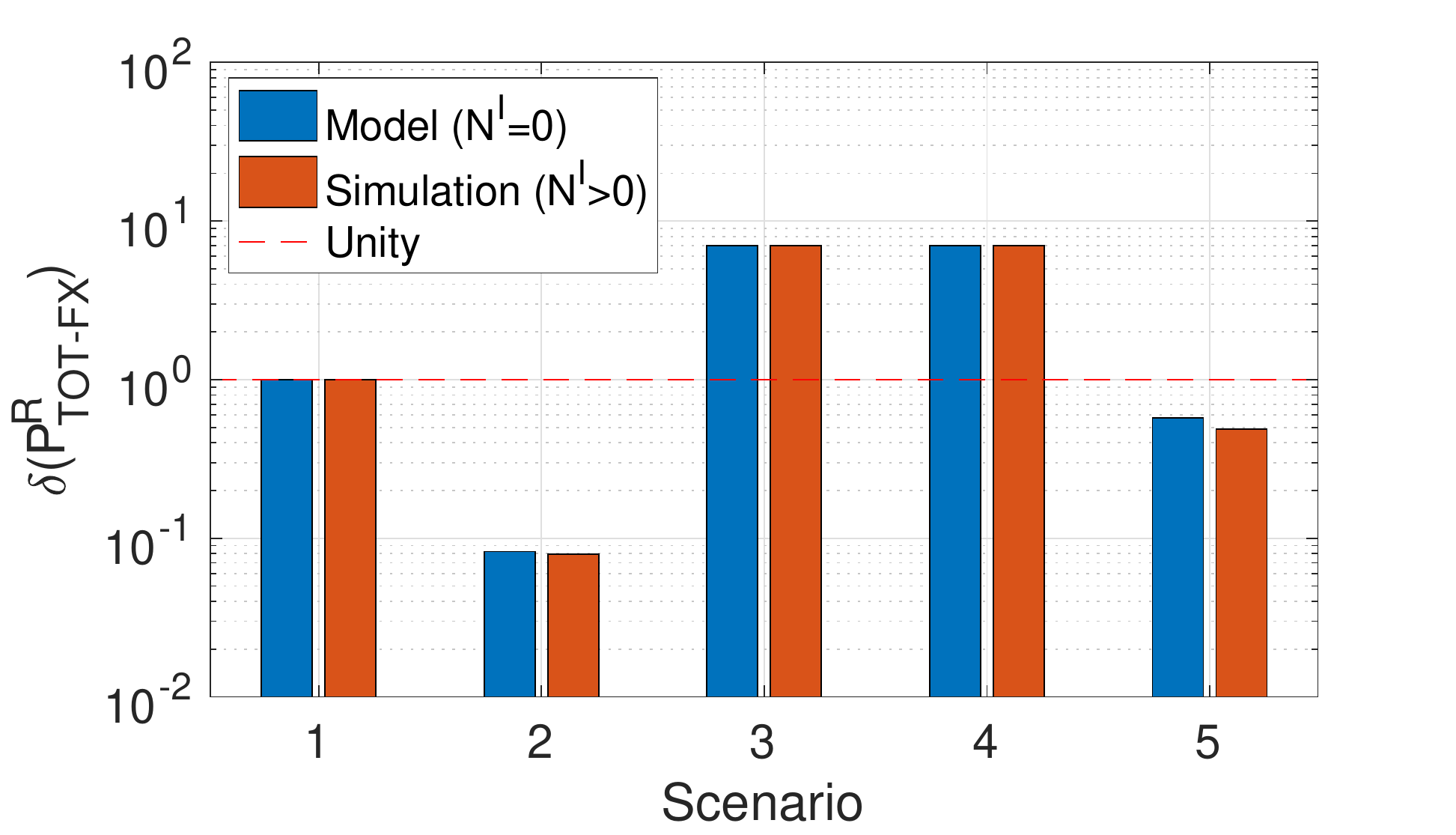}
	\label{fig:SPS_fix}
}
\subfigure[$\delta(P^R_{\text{TOT-CELL}})$]
{
	\includegraphics[width=7cm]{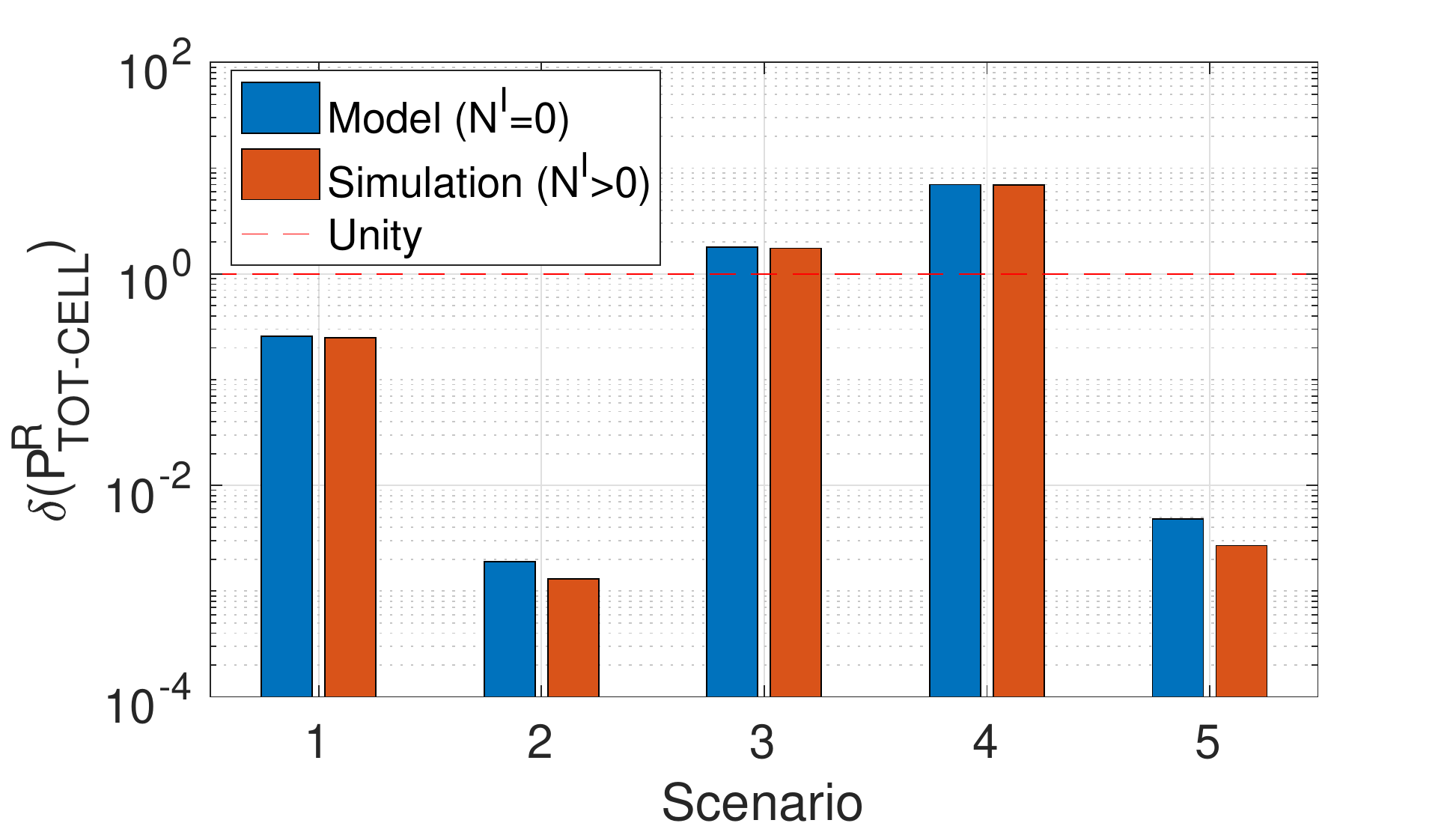}
	\label{fig:SPS_avg}
}
\caption{{Impact of \ac{SPS} policy on the \ac{RFP} ratio at fixed distance $\delta(P^R_{\text{TOT-FX}})$ and on average over the cell $\delta(P^R_{\text{TOT-CELL}})$ over the different scenarios.}}
\label{fig:SPS}
\vspace{-5mm}
\end{figure}

{We initially compute the closed form expressions of $\delta(P^R_{\text{TOT-FX}})$ and $\delta(P^R_{\text{TOT-CELL}})$ when $N^I=0$ (not reported here due to the lack of space). In brief, the expressions of \ac{RFP} are exactly the same of the \ac{ELP} policy for S1 and S2 (since $P^E(1)=P^E(2)$). Therefore, for such scenarios, the same considerations already made for the \ac{ELP} setting hold here. For S3-S5 (i.e., when $f$ varies across the deployment), the closed-form expressions of \ac{RFP} can be derived from the \ac{ELP} ones by multiplying the terms of Tab.~\ref{tab:received_power_ratio_constant} for $\delta(B)=B(1)/B(2)$.}

{In the following step, numerically evaluate the \ac{SPS} policy by imposing $B(1)=20$~[MHz] when $f(1)=700$~[MHz] and $B(2)=3700$~[MHz] when $f(2)=3700$~[MHz], in accordance with the 5G spectrum allocation currently enforced in Italy. Fig.~\ref{fig:SPS} reports $\delta(P^R_{\text{TOT-FX}})$ and $\delta(P^R_{\text{TOT-CELL}})$ over the different scenarios. As expected, the \ac{RFP} ratios of \ac{SPS} are the same of \ac{ELP} in both S1 and S2. Interestingly, a decrease of RFP in deployment (2) w.r.t. (1) is experienced in both S3 and S4 (at both fixed distance and on average over the entire cell). In particular, densification is able to reduce both $\delta (P^R_{\text{TOT-FX}})$ and $\delta (P^R_{\text{TOT-CELL}})$ in S3. Finally, an increase of RFP in deployment (2) is observed in scenario S5 (in line with \ac{ELP} - when $\delta(P^R_{\text{TOT-CELL}})$ is considered).}

\subsection{{Impact of pollution from neighbors}}

{In the last part of our work, we have shed light on the impact of pollution of neighbors, by including the contributions from the second level neighbors w.r.t. the serving \ac{gNB}. Due to the lack of space, we summarize the main outcomes. In particular, we have computed the \ac{RFP} from neighbors as the contributions from the six adjacent \acp{gNB} w.r.t. the serving one (i.e., the first level neighbors) plus the \ac{RFP} generated by the \acp{gNB} that are adjacent to the six neighbors (i.e, the second level neighbors). Results, obtained from both model and simulation across scenarios S1-S5 (\ac{MSP} and \ac{ELP} settings), reveal that the \ac{RFP} is not highly affected by the second level neighbors. Therefore, we can conclude that the \ac{RFP} is mainly impacted by the serving cell and the first level neighbors.}


\section{Summary and Future Works}
\label{sec:conclusions}

We have analyzed the impact of  {{cellular network}} densification on the \ac{RFP}. Initially, we have proposed a simple model to compute the \ac{RFP} from the serving \ac{gNB} \textit{and} a set of neighboring \acp{gNB}. In the following step, we have introduced the \ac{RFP} ratios to compare the ``pollution'' variation among two distinct candidate 5G deployments. Results, obtained by solving the closed-form expressions for the \ac{RFP} ratios in a set of meaningful 5G scenarios, prove that densification \textit{does not} introduce an uncontrolled and exponential increase of \ac{RFP}, thus dispelling this popular myth among the general public. On the contrary, we have demonstrated that densification strongly reduces the \ac{RFP} {(up to three orders of magnitude)} when the radiated power is set according to \ac{MSP}. On the other hand, when the radiated power is set in accordance with \ac{ELP}, the \ac{RFP} variation is always controlled. In this case, the \ac{RFP} increase or decrease depends on the specific densification scenario under consideration. However, there are conditions under which the \ac{RFP} is decreased by densification with \ac{ELP}, e.g., when the frequency is increased in parallel to a light densification, without a change in the propagation exponents. Eventually, we have shown that the outcomes from the model are always in accordance w.r.t. the ones derived by simulation. {Finally, we have analyzed the impact of different key parameters on the \ac{RFP} (e.g., size of exclusion zone, alternative spectrum-based power policies to set the radiated power and increase of pollution from neighbors).}

{As future research activity}, the adoption of detailed propagation models could be an interesting step, in order to consider the impact of shadowing/fading margins and changes in the propagation exponent across the extent of the cell. In addition, the evaluation of indoor 5G deployments adopting mm-Waves frequencies and {femto} cells is another attractive research direction. {Eventually, we will consider the impact of non-regular deployments and/or coverage layouts on the \ac{RFP}}. Finally, the assessment of the \ac{RFP} {by adopting} other exposure metrics, e.g., field strength and/or \ac{PD}, is a promising future {work}.   

\section*{Acknowledgements}
We are grateful to Prof. Andrea Detti, {the handling editor Prof.~Kountouris and the anonymous reviewers for their} fruitful suggestions on how to improve the work. 

\bibliographystyle{ieeetr}

\begin{thebibliography}{10}

\bibitem{chiaraviglio2020will}
L.~Chiaraviglio, G.~Bianchi, N.~Blefari-Melazzi, and M.~Fiore, ``{Will the
  Proliferation of 5G Base Stations Increase the Radio-Frequency
  “Pollution”?},'' in {\em IEEE VTC2020-Spring}, pp.~1--7, IEEE, 2020.

\bibitem{healthconcerns}
{\em {5G appeal: Scientists and doctors warn of potential serious health
  effects of 5G}}.
\newblock Available at
  \url{https://www.jrseco.com/wp-content/uploads/2017-09-13-Scientist-Appeal-5G-Moratorium.pdf},
  last accessed on 25th Sept. 2020.

\bibitem{simko20195g}
M.~Simk{\'o} and M.-O. Mattsson, ``{5G wireless communication and health
  effects - a pragmatic review based on available studies regarding 6 to 100
  GHz},'' {\em Int. journal of environmental research and public health},
  vol.~16, no.~18, p.~3406, 2019.

\bibitem{bushberg2020ieee}
J.~Bushberg, C.~Chou, K.~Foster, R.~Kavet, D.~Maxson, R.~Tell, and M.~Ziskin,
  ``{IEEE Committee on Man and Radiation - COMAR Technical Information
  Statement: Health and Safety Issues Concerning Exposure of the General Public
  to Electromagnetic Energy from 5G Wireless Communications Networks},'' {\em
  Health Physics}, vol.~119, no.~2, p.~236, 2020.

\bibitem{5Grisks}
{\em {Red sky at night, cell tower's alight: nearly half of UK consumers think
  5G is a health risk}}.
\newblock Available at:
  \url{https://tinyurl.com/9crxx8uh},
  Last Accessed: 25th Sept. 2020.

\bibitem{sabotage3}
{\em {The Deep Conspiracy Roots of Europe’s Strange Wave of Cell-Tower
  Fires}}.
\newblock Available at
  \url{https://www.politico.com/news/magazine/2020/05/18/deep-conspiracy-roots-europe-wave-cell-tower-fires-264997},
  Last Accessed: 25th Sept. 2020.

\bibitem{pngban}
{\em {Papua New Guinea govt puts hold on 5G development}}.
\newblock Available at
  \url{https://www.rnz.co.nz/international/pacific-news/406411/png-govt-puts-hold-on-5g-development},
  Last Accessed: 25th Sept. 2020.

\bibitem{jaimaicaban}
{\em {Jamaica must explore risks before adopting 5G tech, Robinson urges}}.
\newblock Available at
  \url{http://www.jamaicaobserver.com/news/jamaica-must-explore-risks-before-adopting-5g-tech-robinson-urges\_177268},
  Last Accessed: 25th Sept. 2020.

\bibitem{brusselsban}
{\em {Radiation concerns halt Brussels 5G development, for now}}.
\newblock Available at
  \url{https://www.brusselstimes.com/brussels/55052/radiation-concerns-halt-brussels-5g-for-now/},
  Last Accessed: 25th Sept. 2020.

\bibitem{californiaban}
{\em {Cities Are Saying No to 5G, Citing Health, Aesthetics - and FCC
  Bullying}}.
\newblock Available at:
  \url{https://tinyurl.com/e8td6t9f},
  Last Accessed: 25th Sept. 2020.

\bibitem{increasedenergy}
{\em {Before it’s too late}}.
\newblock Available at:
  \url{https://www.recorder.com/my-turn-frey-before-it-s-too-late-36084135},
  Last Accessed: 25th Sept. 2020.

\bibitem{chiaraviglio2020health}
L.~Chiaraviglio, A.~Elzanaty, and M.-S. Alouini, ``Health risks associated with
  5g exposure: A view from the communications engineering perspective,'' {\em
  arXiv preprint arXiv:2006.00944}, 2020.

\bibitem{chiaraviglio2021pencil}
L.~Chiaraviglio, S.~Rossetti, S.~Saida, S.~Bartoletti, and N.~Blefari-Melazzi,
  ````pencil beamforming increases human exposure to electromagnetic fields'':
  True or false?,'' {\em IEEE Access}, vol.~9, pp.~25158--25171, 2021.

\bibitem{thurfjell2015network}
M.~Thurfjell, M.~Ericsson, and P.~de~Bruin, ``Network densification impact on
  system capacity,'' in {\em 2015 IEEE 81st vehicular technology conference
  (VTC Spring)}, pp.~1--5, IEEE, 2015.

\bibitem{kountouris2017performance}
V.~M. Nguyen and M.~Kountouris, ``Performance limits of network
  densification,'' {\em IEEE Journal on Selected Areas in Communications},
  vol.~35, no.~6, pp.~1294--1308, 2017.

\bibitem{liu2017network}
J.~Liu, M.~Sheng, L.~Liu, and J.~Li, ``{Network densification in 5G: From the
  short-range communications perspective},'' {\em IEEE Communications
  Magazine}, vol.~55, no.~12, pp.~96--102, 2017.

\bibitem{ge20165g}
X.~Ge, S.~Tu, G.~Mao, C.-X. Wang, and T.~Han, ``{5G ultra-dense cellular
  networks},'' {\em IEEE Wireless Communications}, vol.~23, no.~1, pp.~72--79,
  2016.

\bibitem{park2014asymptotic}
J.~Park, S.-L. Kim, and J.~Zander, ``Asymptotic behavior of ultra-dense
  cellular networks and its economic impact,'' in {\em 2014 IEEE Global
  Communications Conference}, pp.~4941--4946, IEEE, 2014.

\bibitem{arshad2016handover}
R.~Arshad, H.~ElSawy, S.~Sorour, T.~Y. Al-Naffouri, and M.-S. Alouini,
  ``{Handover management in 5G and beyond: A topology aware skipping
  approach},'' {\em IEEE Access}, vol.~4, pp.~9073--9081, 2016.

\bibitem{andrews2016we}
J.~G. Andrews, X.~Zhang, G.~D. Durgin, and A.~K. Gupta, ``Are we approaching
  the fundamental limits of wireless network densification?,'' {\em IEEE
  Communications Magazine}, vol.~54, no.~10, pp.~184--190, 2016.

\bibitem{shafi20175g}
M.~Shafi, A.~F. Molisch, P.~J. Smith, T.~Haustein, P.~Zhu, P.~De~Silva,
  F.~Tufvesson, A.~Benjebbour, and G.~Wunder, ``{5G: A tutorial overview of
  standards, trials, challenges, deployment, and practice},'' {\em IEEE journal
  on selected areas in communications}, vol.~35, no.~6, pp.~1201--1221, 2017.

\bibitem{dang2020should}
S.~Dang, O.~Amin, B.~Shihada, and M.-S. Alouini, ``{What should 6G be?},'' {\em
  Nature Electronics}, vol.~3, no.~1, pp.~20--29, 2020.

\bibitem{oughton2019open}
E.~J. Oughton, K.~Katsaros, F.~Entezami, D.~Kaleshi, and J.~Crowcroft, ``{An
  Open-Source Techno-Economic Assessment Framework for 5G Deployment},'' {\em
  IEEE Access}, vol.~7, pp.~155930--155940, 2019.

\bibitem{chiaraviglio2018planning}
L.~Chiaraviglio, A.~S. Cacciapuoti, G.~Di~Martino, M.~Fiore, M.~Montesano,
  D.~Trucchi, and N.~B. Melazzi, ``{Planning 5G networks under EMF constraints:
  State of the art and vision},'' {\em IEEE Access}, vol.~6, pp.~51021--51037,
  2018.

\bibitem{aaltod3}
{\em D3.1--Study on Small Cells and Dense Cellular Networks Regulatory Issues}.
\newblock Available at
  \url{https://global5g.org/sites/default/files/Global5G.org_D3.1_Study%20on%20small%20cells%20and%20dense%20cellular%20networks%20regulatory%20issues_final.pdf},
  last accessed on 23rd September 2020.

\bibitem{colombi2020analysis}
D.~Colombi, P.~Joshi, B.~Xu, F.~Ghasemifard, V.~Narasaraju, and
  C.~T{\"o}rnevik, ``{Analysis of the Actual Power and EMF Exposure from Base
  Stations in a Commercial 5G Network},'' {\em Applied Sciences}, vol.~10,
  no.~15, p.~5280, 2020.

\bibitem{IEEEC95:19}
``C95.1-2019 {IEEE} standard for safety levels with respect to human exposure
  to radio frequency electromagnetic fields, 3 {kHz} to 300 {GHz},'' tech.
  rep., Institute of Electrical and Electronics Engineers (IEEE), New York, NY,
  US, 2019.

\bibitem{ICNIRPGuidelines:20}
{International Commission on Non-Ionizing Radiation Protection {(ICNIRP)}},
  ``{ICNIRP} guidelines on limiting exposure to time-varying electric, magnetic
  and electromagnetic fields (100 {kHz} to 300 {GHz}).'' Avaiable at:
  \url{https://www.icnirp.org/cms/upload/publications/ICNIRPrfgdl2020.pdf},
  Jul. 2020.
\newblock Last Accessed: 29th Sept. 2020.

\bibitem{3GPPscenarios}
{\em {5G - Study on scenarios and requirements for next generation access
  technologies (3GPP TR 38.913 version 16.0.0 Release 16)}}.
\newblock Available at:
  \url{https://www.etsi.org/deliver/etsi_tr/138900_138999/138913/16.00.00_60/tr_138913v160000p.pdf},
  Last Accessed: 23th March 2021.

\bibitem{rappaport2017overview}
T.~S. Rappaport, Y.~Xing, G.~R. MacCartney, A.~F. Molisch, E.~Mellios, and
  J.~Zhang, ``{Overview of millimeter wave communications for fifth-generation
  (5G) wireless networks with a focus on propagation models},'' {\em IEEE
  Transactions on Antennas and Propagation}, vol.~65, no.~12, pp.~6213--6230,
  2017.

\bibitem{Tho-17}
B.~Thors, A.~Furusk{\"a}r, D.~Colombi, and C.~T{\"o}rnevik, ``{Time-Averaged
  Realistic Maximum Power Levels for the Assessment of Radio Frequency Exposure
  for 5G Radio Base Stations Using Massive MIMO},'' {\em IEEE Access}, vol.~5,
  pp.~19711--19719, 2017.

\bibitem{franci2020experimental}
D.~Franci, S.~Coltellacci, E.~Grillo, S.~Pavoncello, T.~Aureli, R.~Cintoli, and
  M.~D. Migliore, ``An experimental investigation on the impact of duplexing
  and beamforming techniques in field measurements of 5g signals,'' {\em
  Electronics}, vol.~9, no.~2, p.~223, 2020.

\bibitem{adda2020methodology}
S.~Adda, T.~Aureli, S.~Coltellacci, S.~D’Elia, D.~Franci, E.~Grillo,
  N.~Pasquino, S.~Pavoncello, R.~Suman, and M.~Vaccarono, ``A methodology to
  characterize power control systems for limiting exposure to electromagnetic
  fields generated by massive mimo antennas,'' {\em IEEE Access}, vol.~8,
  pp.~171956--171967, 2020.

\bibitem{jamshed2019survey}
M.~A. Jamshed, F.~Heliot, and T.~W. Brown, ``A survey on electromagnetic risk
  assessment and evaluation mechanism for future wireless communication
  systems,'' {\em IEEE Journal of Electromagnetics, RF and Microwaves in
  Medicine and Biology}, vol.~4, no.~1, pp.~24--36, 2019.

\bibitem{friis1946note}
H.~T. Friis, ``A note on a simple transmission formula,'' {\em Proceedings of
  the IRE}, vol.~34, no.~5, pp.~254--256, 1946.

\bibitem{iturec}
{\em {ITU-T K.52 : Guidance on complying with limits for human exposure to
  electromagnetic fields}}.
\newblock Available at \url{https://www.itu.int/rec/T-REC-K.52/en}, last
  accessed on 25th July 2018.

\bibitem{iecexclusion}
{\em {IEC 62232:2017 Determination of RF field strength, power density and SAR
  in the vicinity of radiocommunication base stations for the purpose of
  evaluating human exposure}}.
\newblock Available at \url{https://webstore.iec.ch/publication/28673}, last
  accessed on 25th Sept. 2020.

\bibitem{minsens}
{\em {User Equipment (UE) radio transmission and reception; Part 1: Range 1
  Standalone (3GPP TS 38.101-1 version 16.6.0 Release 16)}}.
\newblock Available at
  \url{https://www.etsi.org/deliver/etsi_ts/138100_138199/13810101/16.06.00_60/ts_13810101v160600p.pdf},
  last accessed on 23th March 2021.

\bibitem{itutks14}
{\em {ITU-T K Supplement 14 The impact of RF-EMF exposure limits stricter than
  the ICNIRP or IEEE guidelines on 4G and 5G mobile network deployment}}.
\newblock Available at
  \url{https://www.itu.int/rec/dologin_pub.asp?lang=e&id=T-REC-K.Sup14-201909-I!!PDF-E&type=items},
  last accessed on 23th March 2021.

\bibitem{itutk70}
{\em {ITU-T K.70 Mitigation techniques to limit human exposure to EMFs in the
  vicinity of radiocommunication stations}}.
\newblock Available at \url{https://www.itu.int/rec/T-REC-K.70-201801-I/en},
  last accessed on 26th Feb. 2020.

\bibitem{isitsafe2020}
L.~{Chiaraviglio}, C.~D. {Paolo}, G.~{Bianchi}, and N.~{Blefari-Melazzi}, ``{Is
  It Safe Living in the Vicinity of Cellular Towers? Analysis of Long-Term
  Human EMF Exposure at Population Scale},'' in {\em 2020 IEEE 91st Vehicular
  Technology Conference (VTC2020-Spring)}, pp.~1--7, 2020.

\bibitem{3GPPchannelmodels}
{\em {5G - Study on channel model for frequencies from 0.5 to 100 GHz (3GPP TR
  38.901 version 16.1.0 Release 16)}}.
\newblock Available at:
  \url{https://www.etsi.org/deliver/etsi_tr/138900_138999/138901/16.01.00_60/tr_138901v160100p.pdf},
  Last Accessed: 23th March 2021.

\bibitem{FCCstd}
{\em {CFR 96.41 - General radio requirements}}.
\newblock Available at: \url{https://www.law.cornell.edu/cfr/text/47/96.41},
  Last Accessed: 23th March 2021.

\end{thebibliography}

\begin{IEEEbiography}[{\includegraphics[width=1in,height=1.25in,clip,keepaspectratio]{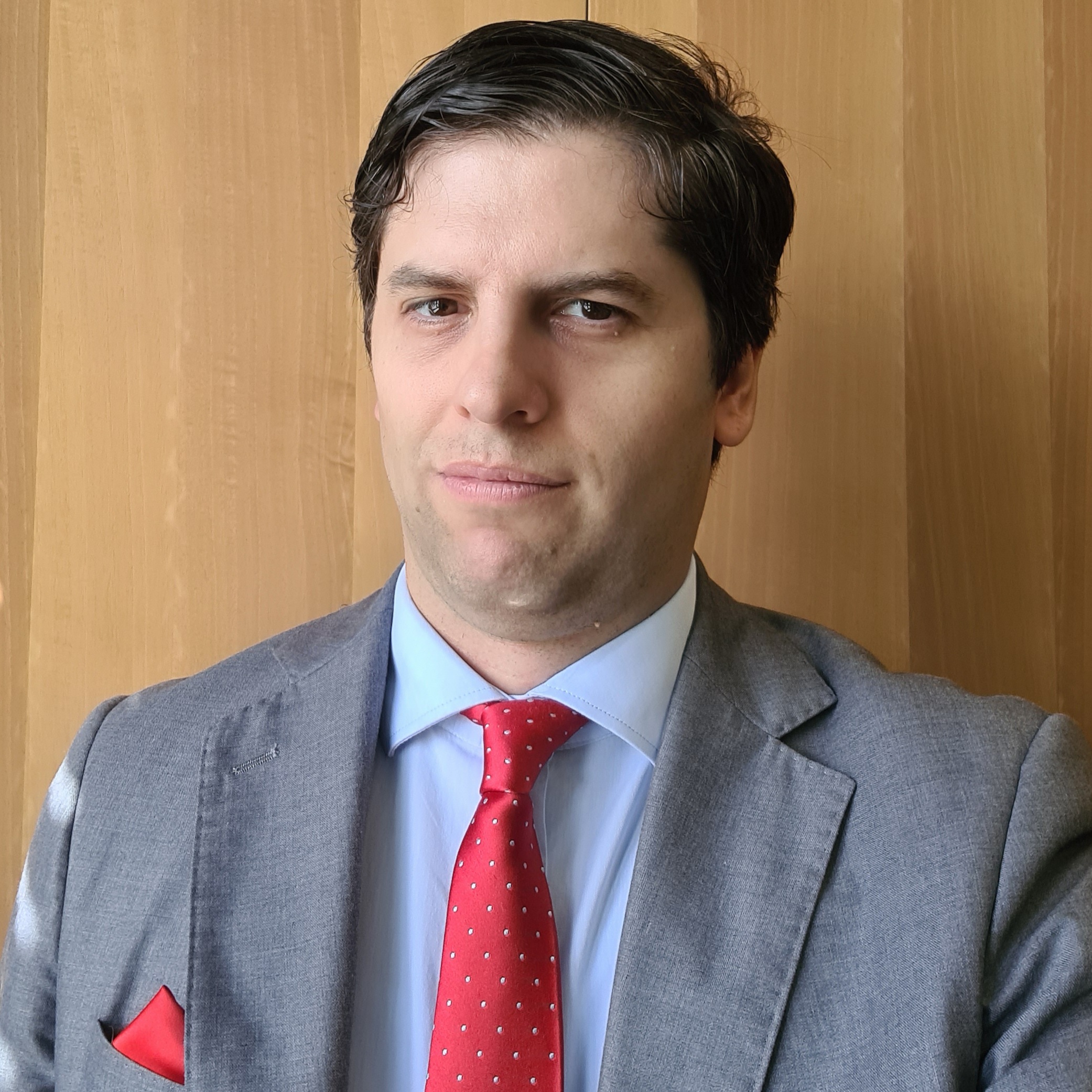}}]{Luca Chiaraviglio} (M'09-SM'16) is Associate Professor at the
University of Rome Tor Vergata (Italy). He holds a Ph.D. in Telecommunication and Electronics Engineering, obtained from Politecnico di Torino (Italy). Luca has co-authored 150+ papers published in international journals, books and conferences. Luca has received the Best Paper Award at IEEE VTC-Spring 2020, IEEE VTC-Spring 2016 and ICIN 2018, all of them appearing as first author. Some of his papers are listed as Best Readings on Green Communications by IEEE. Moreover, he has been recognized as an author in the top 1\% most highly cited papers in the ICT field worldwide. His current research topics cover 5G networks, optimization applied to telecommunication networks, elecromagnetic fields and health risks assessment of 5G communications.
\end{IEEEbiography}

\begin{IEEEbiography}[{\includegraphics[width=1in,height=1.25in,clip,keepaspectratio]{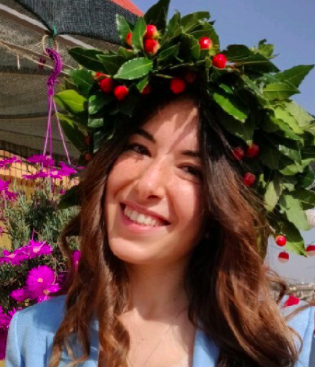}}]{Sara Turco} graduated in ICT and Internet Engineering from the University of Rome Tor Vergata in 2021. Between June 2020 and March 2021 she has been a CNIT Researcher. She currently works as Orizzonte Sistemi Navali S.p.A. as Integration Engineer.
\end{IEEEbiography}

\begin{IEEEbiography}[{\includegraphics[width=1in,height=1.25in,clip,keepaspectratio]{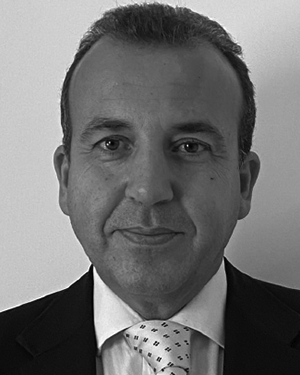}}]{Giuseppe Bianchi} is currently a full professor at the Networking and Network Security, University of Roma Tor Vergata. His research interests include wireless networks (his pioneer work on WLAN modelling received the 2017 ACM SigMobile Test-of-Time Award), programmable network systems, security monitoring and vulnerability assessment, traffic modelling and control, and is documented in about 280 peer-reviewed international journal and conference papers, accounting for more than 20.000 citations (source: Google Scholar). He has coordinated six large scale EU projects, and has been (or still is) editor for several journals in his field, including the IEEE/ACM Transactions on Networking, the IEEE Transactions on Wireless Communications, the IEEE Transactions on Network and Service Management, and the Elsevier Computer Communications.
\end{IEEEbiography}

\begin{IEEEbiography}[{\includegraphics[width=1in,height=1.25in,clip,keepaspectratio]{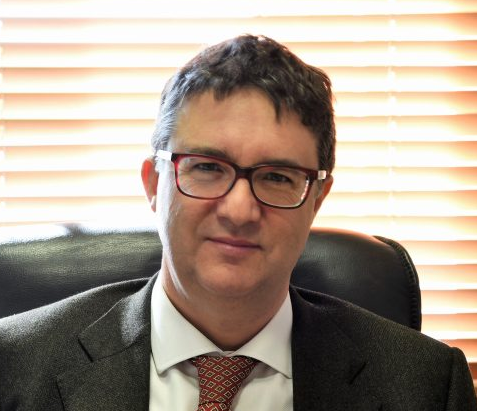}}]{Nicola Blefari-Melazzi} is currently a Full
Professor of telecommunications with the University
of Rome ``Tor Vergata'', Italy. He is currently
the Director of CNIT, a consortium of 37 Italian
Universities. He has participated in over 30 international
projects, and has been the principal investigator
of several EU funded projects. He has been
an Evaluator for many research proposals and a
Reviewer for numerous EU projects. He is the
author/coauthor of about 200 articles, in international
journals and conference proceedings. His research interests include
the performance evaluation, design and control of broadband integrated
networks, wireless LANs, satellite networks, and of the Internet.
\end{IEEEbiography}

\end{document}